\documentclass[prd,aps,amsmath,reprint,nofootinbib,superscriptaddress,floatfix,onecolumn]{revtex4-2} 

\usepackage[normalem]{ulem}

\usepackage{amsfonts}
\usepackage{amssymb}
\usepackage{amsthm}
\usepackage{adjustbox}
\usepackage{xcolor}
\usepackage{axodraw4j}

\usepackage{enumitem}
\usepackage[normalem]{ulem}

\usepackage{graphicx}
\usepackage{subcaption}

\def\prt{\partial}

\def\rf#1{(\ref{#1})}

\newcommand{\beq}{\begin{equation}}
\newcommand{\eeq}{\end{equation}}
\newcommand{\bal}{\begin{aligned}}
\newcommand{\eal}{\end{aligned}}

\begin{document}

\title{Stability and dynamics of bumblebee models}

\author{Quentin G.\ Bailey}
\affiliation{Department of Physics and Astronomy, Embry-Riddle Aeronautical University, 3700 Willow Creek Road, Prescott, AZ, USA}

\author{Robertus Potting}
\affiliation{Departamento de F\'\i sica, Faculdade de Ci\^encias e Tecnologia, Universidade do Algarve, 8005-139 Faro, Portugal}
\affiliation{Centre for Research and Development in Mathematics and Applications (CIDMA), Universidade de Aveiro, Campus de Santiago, 3810-193 Aveiro, Portugal}

\author{Dario T.\ Walter-Cardona}
\affiliation{Department of Physics and Astronomy, Embry-Riddle Aeronautical University, 3700 Willow Creek Road, Prescott, AZ, USA}

\date{\today}

\begin{abstract}
In this work, we study vector field theories with spontaneous Lorentz violation induced by a nonderivative potential \(V(X)\) (with \(X = B_\mu B^\mu\)), known as bumblebee models, with a Maxwell kinetic term in Minkowski spacetime.
We confirm a recent result showing that a class of smooth potentials, including a purely cubic potential, leads to a Hamiltonian density that is bounded from below and assumes a global minimum for nonzero timelike expectation values of the vector field satisfying \(X = b^2 > 0\), inducing spontaneous Lorentz-symmetry breaking.
We show explicitly that the standard quadratic potential has a Hamiltonian that is unbounded from below and thus unstable. 
We analyze the Hamiltonian equations of motion for the timelike case and demonstrate that they are regular at the global minimum.
Furthermore, we examine the linearized spectrum of small fluctuations around the minimum.
We also study particular solutions with spherical symmetry analytically and numerically, 
without assuming linearization.
We find that the time evolution of smooth initial configurations can develop shock-like singularities, and we explain why this behavior is expected in this model.
We also identify a subset of the space of field configurations in which the field equations cease to be hyperbolic, leading to a breakdown of well-posedness of the Cauchy initial-value problem.

\end{abstract}

\maketitle

\section{Introduction}

Recently, there has been a great deal of interest in vector and other models of spontaneous Lorentz symmetry breaking.
The proposal for a bumblebee model for a vector $B_\mu$ with a Maxwell kinetic term and a smooth potential term $V(X)$ (with $X = B_\alpha B^\alpha$), 
driving a vacuum expectation value $\langle B_\mu \rangle = b_\mu$,
dates back to Kosteleck\'y and Samuel \cite{ks89}.
This work was followed by many others
(see, e.g.,
\cite{Kostelecky:2003fs,Bluhm:2004ep,Bailey:2006fd,Chkareuli:2007da})
including one with more general kinetic term, a Lagrange-multiplier potential and couplings to spacetime curvature
\cite{Jacobson:2000xp,Jacobson:2004ts,Deserfest,sarbach19}.
Early studies regarding the stability of bumblebee and other vector models include
Refs.\ \cite{Clayton:2001vy,Elliott:2005va,Eling:2005zq,
Seifert:2007fr,Bluhm:2008yt,Carroll2009,Donnelly:2010qd}.
Later, 
additional studies of stability and quantization were published
\cite{hernaski14,bonder16,assuncao17,belchior23}.
In addition, 
these models have been studied as toy models for producing gravitational and other terms in a general effective field theory framework for testing spacetime symmetries \cite{kl01,Kostelecky:2003fs,Bailey:2006fd,kt11}.

More recently, 
the discovery of an exact Schwarzschild-type black hole solution in gravity coupled to a bumblebee field with a spacelike radial vacuum expectation value of the bumblebee field \cite{Casana:2017jkc} inspired a flurry of new interest in bumblebee models and their solutions;
see, for instance, Refs.\ \cite{Ovgun:2018xys,Maluf:2020kgf,Li:2020dln,Kanzi:2019gtu,Ding:2019mal,Xu:2022frb,Liu:2022dcn}.
In addition, 
the bumblebee model with gravity has been put to the test in cosmology
\cite{bruck26}.
Many works, 
including ones by the authors, 
use a quadratic potential for $V(X)$.
It has been known for some time that, 
due the constraint structure
of the $U(1)$-gauge breaking bumblebee model, 
the Hamiltonian for such a potential turns out to be unbounded from below.
Nonetheless much work has proceeded assuming the bumblebee field lies at the minimum of the potential, 
or the issue has been ignored.

In this work, we re-examine the original bumblebee model in flat spacetime with a smooth potential and examine its stability for different choices of potential.
We find that a potential that is strictly cubic in $X - x$, for some constant $x$, yields a Hamiltonian density  that is strictly bounded from below.
Moreover, its absolute minimum is attained for $X = x$, leading to spontaneous Lorentz-symmetry breaking, whenever the latter is time-like (case $x > 0$).
We also show that it is one of a class of models with this property.
Our results are consistent with those obtained in the recent paper \cite{Zhu:2026hxm}.
We provide more insight by running simulations of time and space-dependent behavior of the vector field.
Most notably, 
these simulations yield evidence of interesting shock-wave-like behavior
worthy of further study.
We also provide a rigorous demonstration that the usually considered quadratic-type potential is indeed unstable at $X = x$.

The paper is organized as follows.
In Section \ref{sec:model} we study the Hamiltonian structure of the model in detail, including aspects of different potential choices for the bumblebee field.
A detailed study of the instability of the quadratic potential in undertaken in Section \ref{quadraticstudy}.
In Section \ref{QFT}, 
details of the model with the cubic potential are studied while expanding around the vacuum value of the bumblebee field, 
including the momentum space solutions, 
the propagator, 
and aspects of quantum interactions.
We return the classical limit in Section \ref{solutions of the classical field equations} and study spherically symmetric static solutions, 
other special cases with simplifications like time-dependence only and the bumblebee lying along one spatial direction.
Most notably in this section we study
the evolution of the bumblebee field using spherical symmetry and encounter interesting behavior.
Finally we discuss and summarize our work in Section \ref{discussion}.

Our conventions include adopting the spacetime metric signature $(+,-, -, -)$, which we note differs from the one commonly adopted for gravity-related studies of the bumblebee.
We use natural units in this paper, $\hbar=1=c$.
However, 
for some of the plots in this work
dimensional quantities like the fields $B_\mu$
or the radial coordinate $r$,
units are chosen to make
them dimensionless.

\section{The model: Lagrangian, Hamiltonian, and Potentials}
\label{sec:model}

We consider the bumblebee model in Minkowski spacetime with Maxwell-type kinetic term:
\begin{equation}
\mathcal{L} = -\frac{1}{4}B_{\mu\nu}B^{\mu\nu} - V(X),
\label{Lagrangian}
\end{equation}
where
\begin{align}
B_{\mu\nu} &= \partial_\mu B_\nu -\partial_\nu B_\mu, \\
X &= B_\mu B^\mu.
\end{align}
In \cite{Bluhm:2008yt,Carroll2009,seifert19,Escobar:2022lpu} a Dirac-Hamiltonian constraint analysis was performed for this model,
which we will first briefly review.
First we display the time derivatives of the Lagrangian density \eqref{Lagrangian} as
\begin{equation}
\mathcal{L} = -\frac{1}{2}(\dot B_i - \partial_i B_0)(\dot B_i - \partial_i B_0) -
\frac{1}{4} B_{ij}B^{ij} - V(X),
\label{Lagrangian2}
\end{equation}
where we use the convention that repeated spatial indices are summed over.
From this we extract the canonical momenta,
\begin{equation}
\pi_i = \dot{B}_i - \partial_i B_0,
\label{pi_i}
\end{equation}
as well as the primary constraint
\begin{equation}
\Omega_1 = \pi_0 \approx 0.
\label{Omega-1}
\end{equation}
where ``$\approx$'' indicates weak equality, i.e., equality on the surface defined by the constraints.
The canonical Hamiltonian density becomes
\begin{align}
\mathcal{H}_c &= \pi_i \dot{B}_i - \mathcal{L} \nonumber\\
&= \frac{1}{2}\pi_i \pi_i - B_0\partial_i\pi_i + \frac{1}{4}B_{ij}B_{ij} + V(X),
\end{align}
where we performed a partial integration in the second term since $H=\int d^3x {\cal H}$.

The primary Hamiltonian density is defined
as
\begin{equation}
\mathcal{H}_p = \mathcal{H}_c + \mu\Omega_1,
\label{primary-Hamiltonian}
\end{equation}
where $\mu$ is a function of the phase-space variables that is to be determined.
Next we have to impose that the primary constraint is conserved in time,
which yields a secondary constraint:
\begin{equation}
\dot{\Omega}_1 = \{\pi_0, H_p\} = \partial_i \pi_i - V'(X)2B_0 \equiv \Omega_2 \approx 0.
\label{Omega_2}
\end{equation}
The Poisson bracket of $\Omega_1$ and $\Omega_2$ is nonzero:
\begin{equation}
\{\Omega_1(x),\Omega_2(y)\} = 2\bigl(V'(X) + 2B_0^2 V''(X)\bigr)\delta(x-y),
\end{equation}
so $\Omega_1$ and $\Omega_2$ form a second-class system.
For the time evolution of $\Omega_2$ it follows:
\begin{equation}
\dot{\Omega}_2 = \{\Omega_2,H_p\} 
\approx 2\partial_i(V' B_i) + 4B_0B_i V''(\pi_i + \partial_i B_0) - 2\mu(V' + 2B_0^2 V''),
\end{equation}
where we have dropped the $X$ dependence of $V'$ and $V''$, for notational simplicity.
Imposing that this vanish weakly implies that
\begin{equation}
\mu \approx \frac{\partial_i(V'B_i) + 2B_0B_i V''(\pi_i + \partial_i B_0)}{V' + 2B_0^2 V''}.
\label{mu}
\end{equation}
Thus there are no further constraints.
It follows that the phase space has $8 - 2 = 6$ local degrees of freedom.

Let us next investigate stability.
By using the constraint $\Omega_2$ we can express the Hamiltonian density as
\begin{equation}
\mathcal{H}_p \approx \mathcal{H}_c \approx
\frac{1}{2}\pi_i \pi_i + \frac{1}{4}B_{ij}B_{ij} + V - 2B_0^2V'.
\label{H_p}
\end{equation}
Defining
\begin{equation}
\mathcal{H}_V = V - 2B_0^2V',
\label{H_V}
\end{equation}
we see that 
if $\mathcal{H}_V$ is bounded from below, the same is true for $\mathcal{H}_p$.
From the definition \eqref{H_V}
we see that a necessary (though not sufficient) condition for this to be the case
is that $V'$ be negative semi-definite.
Indeed, if $V'(X)$ is positive for any value of $X$,
$\mathcal{H}_V$ can be made arbitrarily negative as $B_0$ is taken to infinity.

Let us consider some key examples below.
\begin{description}[leftmargin=12pt]
\item[1.\ $V = -\frac{m^2}{2}X$]
This case corresponds to the Proca action describing a massive photon.
$V' = -\frac{m^2}{2} < 0$ is negative definite.
We find  $\mathcal{H}_V = \frac{m^2}{2}(B_0^2 + B_i B_i)$,
which is clearly positive semi-definite.
However, it does not imply any spontaneous Lorentz violation.
\item[2.\ $V = \frac{1}{2}\lambda(X - x)^2$ (with $x$ constant)]
This corresponds to the case usually considered in the literature on the bumblebee.
As $V'$ is not semi-negative definite for any value of $\lambda$,
$H_V$ cannot be bounded from below.
Indeed, one finds that
$\mathcal{H}_V = -\frac{1}{2}\lambda(B_0^2 - B_i B_i - x)(3B_0^2 + B_i B_i + x)$.
This is clearly not positive definite,
as it can always be made arbitrarily negative for any choices
of $\lambda$ and $x$, leading to an unstable system.
The same conclusion holds for $V \propto (X-x)^{2n}$ with $n$ integer.
Plots for the timelike and spacelike cases of ${\cal H}_V$ are provided in Figs.\ \ref{fig:quad-potential}
for $\lambda > 0$.
\item[3.\ $V = -\frac{\lambda}{3}(X - x)^3$]
In this case we have $V' = -\lambda(X - x)^2 \le 0$,
which is negative definite for $\lambda > 0$,
and the effective potential $\mathcal{H}_V = \frac{\lambda}{3} (X - x)^2(5B_0^2 + B_i B_i + x)$.
For the case $x \ge 0$ it follows that $\mathcal{H}_V \ge 0$,
while $\mathcal{H}_V = 0$ iff $X = x$,
corresponding to the two-sheeted hyperboloid $B_0^2 = B_i^2 + x$.
A plot is provided in Fig.\ \ref{fig:plot_HV_timelike}.
The dynamical system is stable and leads to spontaneous Lorentz violation,
with timelike expectation value for the bumblebee field.
For the case $x < 0$,
$\mathcal{H}_V \ge 0$ outside the ellipsoid defined by $5B_0^2 + B_i B_i = -x$,
and $\mathcal{H}_V < 0$ inside, which is a closed domain.
As $\mathcal{H}_V$ is a smooth function, it must be bounded from below.
By shifting it by a suitable constant,
we can arrange it to be semi-positive definite.
A plot is provided in Fig.\ \ref{fig:plot_HV_spacelike} (for the unshifted system).

\begin{figure}[h!]
  \begin{subfigure}[h]{0.48\linewidth}
    \includegraphics[width=\linewidth]{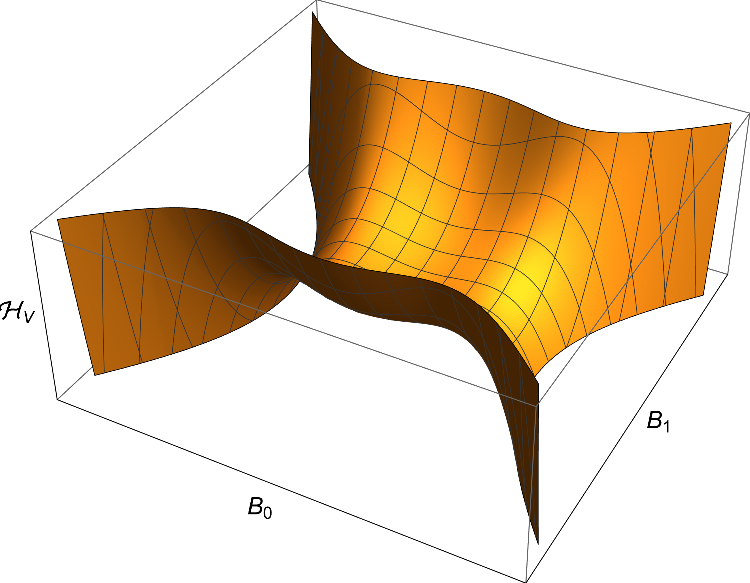}
    \caption{The timelike case $x > 0$ for the quadratic potential.}
    \label{fig:plot_HV_tl_quad}
  \end{subfigure}
 \hfill
  \begin{subfigure}[h]{0.48\linewidth}
    \includegraphics[width=\linewidth]{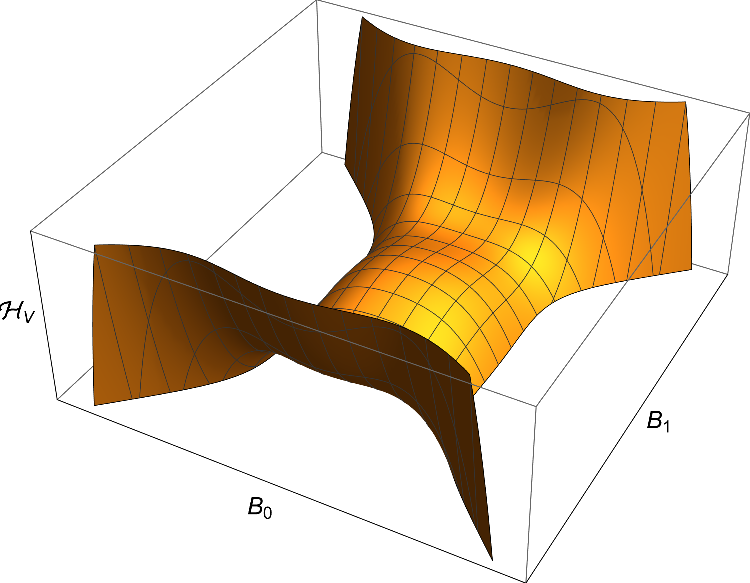}
    \caption{The spacelike case $x < 0$ for the quadratic potential.}
    \label{fig:plot_HV_sl_quad}
  \end{subfigure}%
 \caption{Plots of $\mathcal{H}_V$ for the quadratic potential as a function of $B_0$ and $B_1$, with $B_2 = B_3 = 0$ and $\lambda > 0$.}
 \label{fig:quad-potential}
 \end{figure}

 \begin{figure}[h!]
  \begin{subfigure}[h]{0.48\linewidth}
    \includegraphics[width=\linewidth]{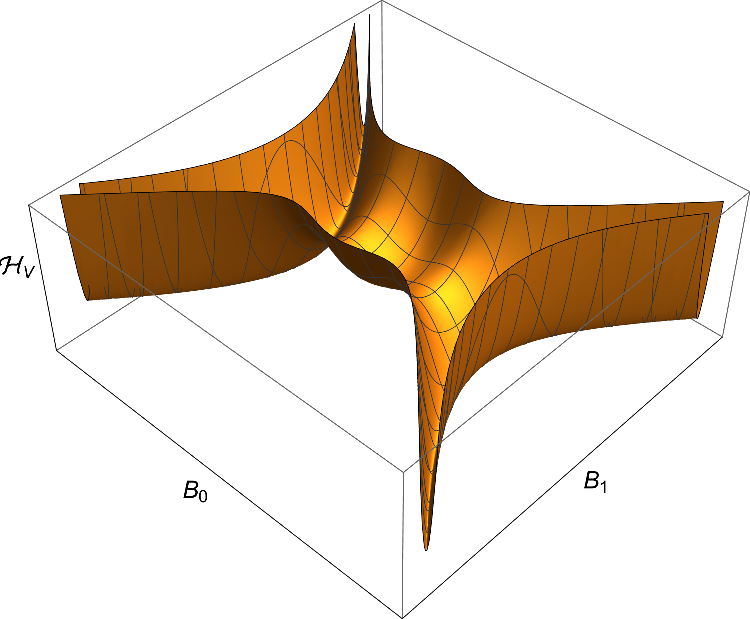}
    \caption{The case $x > 0$, leading to spontaneous lorentz violation
    with timelike expectation value for $B_\mu$.}
    \label{fig:plot_HV_timelike}
  \end{subfigure}
 \hfill
  \begin{subfigure}[h]{0.48\linewidth}
    \includegraphics[width=\linewidth]{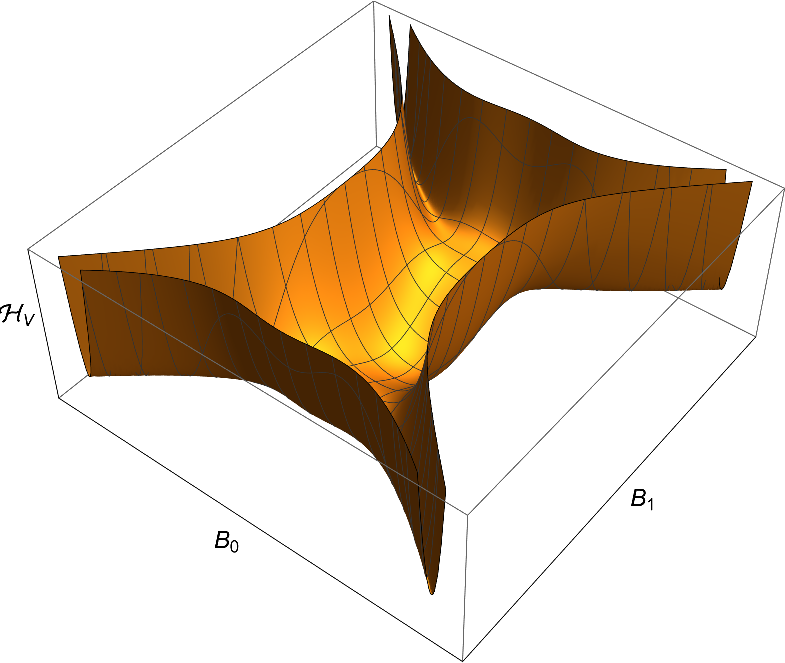}
    \caption{The case $x < 0$. $\mathcal{H}_V$ has its absolute minimum
    at $B_0 = B_1 = 0$, and is stationary for $B_1^2 = B_0^2 - x$.
    }
    \label{fig:plot_HV_spacelike}
  \end{subfigure}%
 \caption{Plots of $\mathcal{H}_V$ for the cubic potential as a function of $B_0$ and $B_1$ (with $B_2 = B_3 = 0$).}
 \label{fig:cubic-potential}
 \end{figure}

\end{description}

We also remark in passing that other potentials for $B_\mu$ have been suggested in the literature.
For example, 
nonpolynomial potentials based on hypergeometric functions were found to solve the renormalization group equation at one-loop for the model in \rf{Lagrangian} \cite{ak05}.
This was investigated for classical gravity solutions in Ref.\ \cite{bmw25}.
In fact, it turns out that the Hamiltonian for this type of potential is unbounded below.
\vspace{3mm}

Let us now investigate the stationary
points of $\mathcal{H}_V$ and their properties. We have
\begin{align}
\frac{\partial \mathcal{H}_V}{\partial B_0} &= -2B_0(V' + 2B_0^2 V''), \\
\frac{\partial \mathcal{H}_V}{\partial B_i} &= -2B_i(V' - 2B_0^2 V'').
\label{extrema}
\end{align}
These conditions lead to 4 types of stationary points:
\begin{enumerate}
\item $B_0 = B_i = 0$: stationary point at which the vector field is identially zero;
\item $B_0 = 0$ and $V' = 0$: stationary points at
nonzero $B_i$, but with $B_0 = 0$ fixed;
\item $B_i = 0$ and $V' + 2B_0^2 V''= 0$: stationary point(s) at nonzero $B_0$,
but with $B_i = 0$ fixed;
\item $V' = V'' = 0$: stationary points at fixed value
of $X$, with no additional conditions on the components $B_\mu$.
\end{enumerate}

Examples of the stationary points of type 1 are provided by any of the potentials considered above by inspecting the figures.
For the Proca action and the spacelike cubic potential it corresponds to the absolute minimum of $\mathcal{H}_V$,
for the timelike quadratic and cubic potentials a local minimum,
while for the spacelike quadratic potential it is a local maximum (considering always $\lambda > 0$).
Of course, as the expectation value of the field is zero, there is no spontaneous Lorentz violation.

Stationary points of type 2 can be identified for the spacelike quadratic and cubic potentials,
and those of type 3 for the timelike versions.
In each of these cases they are saddle-point-like and unstable,
and therefore do not induce spontaneous Lorentz breaking.

Most interesting, for our purposes, are the stationary points of type 4.
They do not exist for the Proca action and the quadratic potential,
but for the cubic potential they do.
In the time-like case they constitute a two-sheeted hyperboloid 
corresponding to the absolute, degenerate minimum of $\mathcal{H}_V$,
which will naturally lead to spontaneous Lorentz breaking.
In the spacelike case they constitute a one-sheeted hyperboloid,
but do not correspond to the absolute minimum of $\mathcal{H}_V$,
which is at zero field value.
The latter can be reached from the stationary points of type 2
located on the hyperboloid at $B_0 = 0$, $B_iB_i = x$ by a continuous
downward path.
Therefore, in the spacelike case the type 4 stationary points are unstable.
\\

We can determine the nature of the stationary points by considering the Hessian of $\mathcal{H}_V$, which is given by
\begin{equation}
M^{\mu\nu} = \frac {\prt^2 {\cal H}_V}{\prt B_\mu \prt B_\nu}.
\label{hessian}
\end{equation}
yielding the components \cite{Zhu:2026hxm}
\begin{equation}
\bal
M^{00} &= -2 V^\prime - 16 B_0^2 V^{\prime \prime} -8 B_0^4 V^{\prime \prime \prime},
\\
M^{0j} &= 4 B_0 B_j (V^{\prime \prime} + 2 B_0^2 V^{\prime \prime \prime}),
\\
M^{jk} &= -2 \delta_{jk} (V^\prime - 2 B_0^2 V^{\prime \prime}) 
+ 4 B_j B_k ( V^{\prime \prime} - 2 B_0^2 
 V^{ \prime \prime \prime }). 
\label{hessian2}
\eal
\end{equation}

For the most interesting case 4 above,
a stationary point with $V^\prime = 0 = V^{\prime \prime}$, 
it follows that Eqs.\ \eqref{hessian2} reduce to the bi-vector form
\beq
M^{\mu\nu} = -8B_0^2 V''' B^\mu B^\nu.
\eeq
It has one nonzero eigenvector $X_\mu = \bar{B}_\mu \equiv B^\mu $
with eigenvalue 
\begin{equation}
    -8B_0^2(B\cdot \bar{B}) V''' = -8B_0^2(B_0^2 + \vec{B}^2)V'''.
\end{equation}
as well as a triply degenerate eigenvalue zero, corresponding
to eigenvectors $X_\mu$ satisfying $B\cdot X = 0$.
Of course, the latter correspond to the directions parallel to the
three-dimensional submanifold (hyperboloid) defined by the minimum of $\mathcal{H}_V$,
while the nonzero eigenvalue correspond to the perpendicular direction.
We see that, whenever $V''' < 0$ as well as $B_0\ne 0$ at the stationary point, it corresponds to a (degenerate) local minimum.
For general potentials, whether it also corresponds to a global minimum remains to be checked separately.
As we have seen above, this is the case for the purely cubic potential with $x > 0$ (i.e., timelike expectation value) (see also Ref.\ \cite{Zhu:2026hxm}), while in the purely cubic case with $x < 0$, $\mathcal{H}_V$ is bounded from below but with its global minimum at $B_0 = B_i = 0$.

For general choices of $V(X)$, it is surprisingly tricky to determine whether or not $\mathcal{H}_V$ is bounded from below and whether it assumes its global minimum at $X = x$.
However, it was demonstrated in Ref.\ \cite{Zhu:2026hxm} that there are a few general criteria that apply (which we translated to our time-like metric convention):
\begin{enumerate}
     \item Already mentioned above, a necessary condition for $\mathcal{H}_V$ to be bounded from below is $V'(X) \le 0$.
    \item $\mathcal{H}_V$ can only assume its global minimum for $X = x$ if $x > 0$, so that the bumblebee acquires a timelike vacuum value.
    If $x < 0$, the degenerate stationary point does not correspond to a global minimum of $\mathcal{H}_V$.
    \item Expressing $V(X) = (X - x)^3f(X)$, with $f(X)$ some smooth function and $x > 0$,
    a sufficient condition for $\mathcal{H}_V$ to be bounded below with global minimum at $X = x$ is that $f(X) < 0$ as well as $(X - x)f'(X) \le 0$.
\end{enumerate}
Examples of potentials satisfying these conditions are the purely cubic potential $V = -\frac{\lambda}{3}(X - x)^3$ ($\lambda > 0$), as well as
any expression of the type
\begin{equation}
V(X) = a_3(X-x)^3 + a_5(X-x)^5 + a_7(X-x)^7 + \ldots
\label{V-odd}
\end{equation}
with $a_{2i+1} \le 0$.
These include polynomial potentials
(if there is a maximum value for $n$ for which $a_{2n-1}$ is nonzero)
or nonpolynomial ones.

\begin{figure}[h!]
 \begin{subfigure}[h]{0.48\linewidth}
   \includegraphics[width=\linewidth]{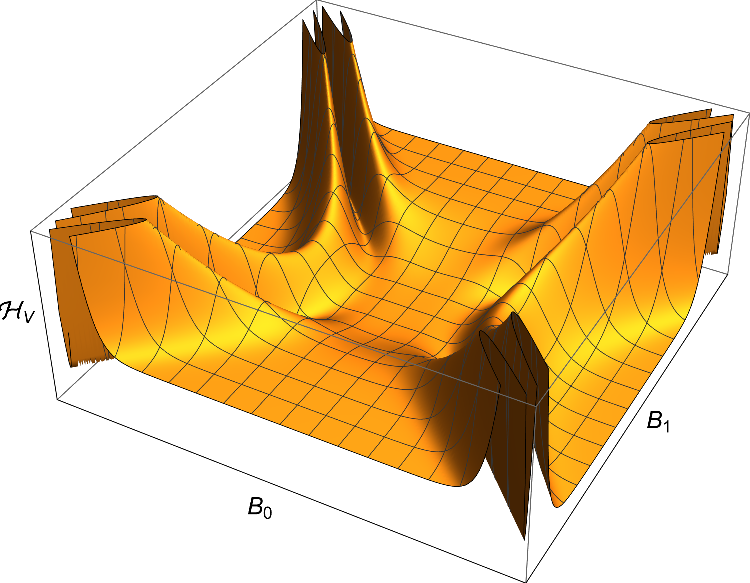}
   \caption{3D plot of $\mathcal{H}_V$ as a function of $B_0$ and $B_1$, taking $B_2 = B_3 = 0$, $\lambda = 0.25$ and $x = 1$. Note the degenerate local minimum at the two-sheeted hyperboloid defined by $X = x$.}
   \label{fig:tanh-cubed-potential}
 \end{subfigure}
\hfill
 \begin{subfigure}[h]{0.42\linewidth}
   \includegraphics[width=\linewidth]{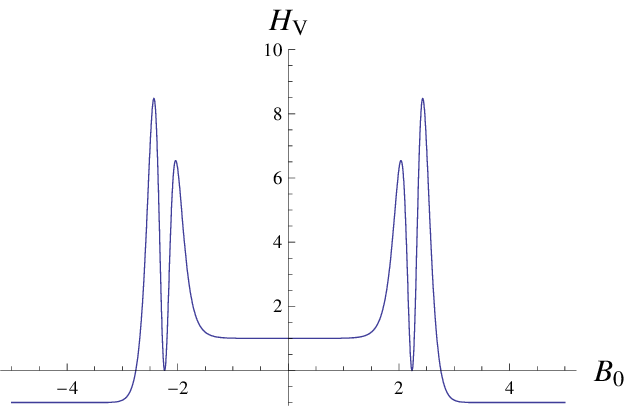}
   \caption{Plot of $\mathcal{H}_V$ as a function of $B_0$ for fixed $B_1 = 3$, $B_2 = B_3 = 0$. Note the occurrence of local minima between steep peaks when $X = x$.}
   \label{fig:tanh-cubed-fixed-B1}
 \end{subfigure}%
\caption{Plots representing the scalar potential $\mathcal{H}_V$ for the tanh-cubed potential of Eq.\ \ref{tanh-cubed-potential}.}
\label{fig:tanh-cubed}
\end{figure}

For an example of a non-polynomial potential with desirable properties consider
\begin{equation}
    V(X) = - V_0\left[\tanh\bigl(\lambda(X - x)\bigr)\right]^3
    \label{tanh-cubed-potential}
\end{equation}
It follows from \rf{H_V} that
\begin{equation}
\mathcal{H}_V = \frac{V_0}{2}\left[\frac{\tanh\bigl(\lambda(X - x)\bigr)}{\cosh\bigl(\lambda(X - x)\bigr)}\right]^2 \left[12\lambda B_0^2 - \sinh\bigl(2\lambda(X - x)\bigr) \right].
\end{equation}
The functional dependence of $\mathcal{H}_V$ as a function of the components of the vector field is represented in Fig.\ \ref{fig:tanh-cubed}.

We see from the figures \ref{fig:tanh-cubed-potential} and \ref{fig:tanh-cubed-fixed-B1} that it is bounded from below and that there is a stable local minimum at $X = x$, that can induce spontaneous Lorentz breaking.
Interestingly, however, the absolute minimum of the potential (at $-V_0$) does not occur at $X = x$, but for $B_1 \to \infty$, $B_0$ fixed. 
Also, there is a plateau in between the sheets of the hyperboloid on which $\mathcal{H} \approx V_0$.
These properties are consistent with the fact that this potential satisfies criterion 1 above, but not criterion 3.

\section{Demonstration that the Hamiltonian of the quadratic potential is unbounded below}
\label{quadraticstudy}

We have seen above that the Hamiltonian density $\mathcal{H}_p$ (defined in Eq.\ \eqref{H_p}) is bounded from below if this is the case for $\mathcal{H}_V$. 
However, the converse is not necessarily true, that is,
$\mathcal{H}_V$ being unbounded from below does not necessarily imply
that this is the case for the full Hamiltonian density $\mathcal{H}_p$.
The reason is that $\mathcal{H}_p$ includes a positive definite term $\frac{1}{2}\pi_i \pi_i$
that is not independent of $B_\mu$ due to the (Gauss-like) constraint $\Omega_2$ \rf{Omega_2}.
It is conceivable that, even if $\mathcal{H}_V$ takes negative values, this is compensated by a positive contribution from $\frac{1}{2}\pi_i \pi_i$, and that the full Hamiltonian remains bounded from below.

An important case for which $\mathcal{H}_V$ is unbounded from below is the quadratic potential, which has been considered in numerous works in the literature.
We will show in this section that in fact its total Hamiltonian is unbounded from below.
We will do this by showing that there exist small perturbations from the vacuum value $X = x$, consistent with the constraints, that can turn the Hamiltonian arbitrarily negative.

We consider the potential
\begin{equation}
V(X) = \frac{\lambda}{2}\bigl(X - b^2\bigr)^2 = \frac{\lambda}{2}(\delta X)^2,
\label{quadratic-V}
\end{equation}
with $X = B_0^2 - B_i^2$
and $\delta X = X - b^2$. 
The Hamiltonian density is
\begin{align}
\mathcal{H} &= \frac{1}{2}\vec{\pi}\/^2 + \frac{1}{4}B_{ij}B_{ij} + V(X) - 2B_0^2V'(X)\nonumber\\
&= \frac{1}{2}\vec{\pi}\/^2 + \frac{1}{4}B_{ij}B_{ij} + \mathcal{H}_V,
\label{Ham-onshell}
\end{align}
with $\cal H_V$ comprising the last two terms, 
as in Eq.\ \rf{H_V}.
Note that $V'(X) = \lambda\,\delta X$.
The canonical momenta or ``electric field" $\vec{\pi}$ is subject to the Gauss constraint
\begin{equation}
\vec{\nabla}\cdot\vec{\pi} = 2B_0V'(X) = 2\lambda B_0(\delta X).
\label{constraint}
\end{equation}

Vacuum configurations are those in which the field sits in the minimum
of the potential $V(X)$, in other words, they satisfy $\delta X = 0$.
We want to demonstrate that the Hamiltonian is unbounded from below.
To show this, we aim to construct a field configuration depending on some 
adjustable constant(s) that can be tuned such that the energy of the
configuration be made arbitrarily negative.
Moreover, we are interested in a configuration that is local
in the sense that it is only different from the vacuum configuration
within a fixed, bounded region of space.

\subsection{Piecewise-constant configuration}

First, we consider the spherically symmetric configuration in which
the bumblebee components take the form
\begin{align}
\label{initial_B0_stepwise}
  B_0(r) &=
  \begin{cases} 
    \bar{B}_0,& \text{for }0\le r < a; \\
    b, & \text{for }a\le r < R; \\
    \tilde{B}_0, & \text{for }R < r < R + \delta R; \\
    b, & \text{for }r > R + \delta R;
   \end{cases}\\
   B_i(r) &= 0.
\end{align}
Here $\bar{B}_0$, $\tilde{B}_0$, $a$, $R$ and $\delta R$ are constants,
where we will consider take $\delta R \ll R$ for convenience.
The values of $\bar{B}_0$ and $\tilde{B}_0$ will be specified later.
It follows that $B_0(r)$ is piecewise constant
(it could be easily smoothened to a continuous function,
which we consider in the next subsection).
It is different from its vacuum value within a sphere of radius $a$
as well as inside a spherical shell with inner radius $R$
and thickness $\delta R$.

From the constraint \eqref{constraint} we see that the expression,
\begin{equation}
2\lambda B_0 \delta X,
\end{equation}
represents an effective charge density.
For our configuration it is nonzero within the sphere of radius $a$,
as well as inside the spherical shell.
In the analysis below,
we will choose the values of $\bar{B}_0$ and $\tilde{B}_0$ such that
the total charge of the sphere plus that of the shell is zero.
Therefore, we have that the total charge of the sphere ($Q$)
is opposite to that of the shell:
\begin{equation}
Q = \frac{4}{3}\pi a^3 2\lambda\bar{B}_0\,\delta\bar{X}
= -\frac{4}{3}\pi\bigl((R + \delta R)^3 - R^3\bigr)2\lambda\tilde{B}_0\,\delta\tilde{X}.
\label{Q}
\end{equation}
Here $\delta\bar{X} = \bar{B}_0^2 - b^2$ and
$\delta\tilde{X} = \tilde{B}_0^2 - b^2$.
Note that $Q$ is a dimensionless parameter which we are free to choose arbitrarily.
For the time being, we will just specify that it is positive.

By Gauss' law, it then follows that the electric field
$\vec\pi$ is radial for $0\le r < R +\delta R$,
and zero for $r > R +\delta R$.
Explicitly, we have
\begin{equation}
\pi_r(r) =
\begin{cases}
\frac{Q\,r}{4\pi a^3},& \text{for } 0 \le r \le a;\\
\frac{Q}{4\pi r^2},& \text{for } a \le r \le R;\\
\frac{Q}{4\pi r^2}\frac{(R + \delta R)^3 - r^3}{(R + \delta R)^3 - R^3},& \text{for } R \le r \le R + \delta R;\\
0, &\text{for } r \ge R+\delta R.
\end{cases}
\end{equation}
For the energy of the electric field, 
which is the kinetic energy term in the full Hamiltonian density \rf{Ham-onshell},
we obtain by integrating over volume,
\begin{align}
\frac{1}{2}\int_0^\infty \pi_r^2 4\pi r^2dr &\approx
2\pi\int_0^a\left(\frac{Q\,r}{4\pi a^3}\right)^2 r^2dr
+ 2\pi\int_a^R\left(\frac{Q}{4\pi r^2}\right)^2 r^2 dr \nonumber\\
&=\frac{Q^2}{8\pi}\left(\frac{6}{5a} - \frac{1}{R}\right),
\label{electric-field-energy}
\end{align}
where we disregarded the $\mathcal{O}(\delta R)$ contribution
to the integral from the shell.

Next we add the contribution to the energy from $\mathcal{H}_V$,
which gives
\begin{align}
\int_0^a + \int_R^{R + \delta R}&\left[\frac{\lambda}{2}(\delta X)^2 - 2B_0^2\lambda\,\delta X\right]4\pi r^2dr\nonumber\\
&= \frac{4\pi}{3}a^3\lambda\left(\frac{1}{2}\delta\bar{X}^2 - 2\bar{B}_0^2\delta\bar{X} \right)
+ 4\pi R^2\delta R\lambda\left(\frac{1}{2}\delta\tilde{X}^2 - 2\bar{B}_0^2\delta\tilde{X} \right).
\label{potential-energy}
\end{align}
Note that $\bar{B}_0$ and $\tilde{B}_0$ satisfy Eqs.\ \eqref{Q}.
In fact, the product $\bar{B}_0\,\delta\bar{X}$ is a cubic expression
in terms of $\bar{B}_0$ of the form
\begin{equation}
\bar{B}_0(\bar{B}_0^2 - b^2) = \alpha^3,
\label{cubic}
\end{equation}
with $\alpha^3 = 3Q/(8\pi a^3\lambda)$,
which has zeros at $\bar{B}_0 = -b, 0$ and $b$.
For any positive value of $\alpha$ (which is the case for $Q > 0$),
Eq.\ \eqref{cubic} has a solution for $\bar{B}_0$ that is larger than $b$.
We will be interested in taking the limit of small $a$,
and thus of large $\alpha$.
In this case, 
the solution to Eq.\ \eqref{cubic} becomes
approximately
\begin{equation}
\bar{B}_0 \approx \alpha = \left(\frac{3Q}{8\pi\lambda}\right)^{1/3}\frac{1}{a},
\end{equation}
while $\delta\bar{X} \approx \alpha^2$.
The first term on the right-hand side of Eq.\ \eqref{potential-energy} then becomes
approximately
\begin{equation}
-2\pi a^3\lambda\alpha^4 = -\frac{3}{4}\left(\frac{3}{8\pi\lambda}\right)^{1/3}  \frac{Q^{4/3}}{a},
\label{potential-energy-divergent}
\end{equation}
which diverges in the small-$a$ limit.

We still have to add the contribution of the second term on  the right-hand side 
of Eq.\ \eqref{potential-energy}.
From Eq.\ \eqref{Q} we see that the product
$\tilde{B}_0\delta\tilde{X}=\tilde{B}_0(\tilde{B}_0^2 - b^2)$ is constrained
to be equal to $-Q/(8\pi R^2\delta R)$.
A solution to Eq.\ \eqref{cubic} exists for positive $\tilde{B}_0$ in the
interval $b/\sqrt{3} < \tilde{B}_0 < b$ if the condition
\begin{equation}
\frac{Q}{8\pi R^2\delta R} < b^3 \frac{2}{\sqrt{27}},
\end{equation}
is satisfied, 
which we will assume to be the case.
This means we have to choose the volume of the shell sufficiently large:
\begin{equation}
R^2\delta R > \frac{\sqrt{27}Q}{16\pi\lambda b^3}.
\label{volume-shell}
\end{equation}
It is important to note that this minimum volume of the shell is independent on the
choice of the radius $a$ of the inner sphere.
The  contribution of the second term on  the right-hand side 
of Eq.\ \eqref{potential-energy} then takes the negative, $a$-independent value
\begin{equation}
-2\pi R^2\delta R \lambda(b^2-\tilde{B}_0^2)(3\tilde{B}_0^2 + b^2).
\label{energy-shell}
\end{equation}

Combining the results obtained above, we find that the sum of the contributions
\eqref{electric-field-energy} and \eqref{potential-energy} contains $a$-dependent
as well as $a$-independent terms.
In the limit of small $a$, the former yield the divergent contributions 
given by Eqs.\ \eqref{electric-field-energy} and \eqref{potential-energy-divergent}.
Both terms diverge as $1/a$, 
but with opposite signs, 
thus
\begin{equation}
H \supset \,  \frac{3Q^{4/3}}{4\pi a}\left(\frac{Q^{2/3}}{5} -\left(\frac{3\pi^2}{8\lambda}\right)^{1/3}\right).
\label{total-divergent}
\end{equation}
This expression is negative if the charge $Q$ is taken to be sufficiently small,
namely if
\begin{equation}
Q < \sqrt{\frac{375}{8\lambda}}\,\pi.
\label{condition_Q}
\end{equation}
It then becomes arbitrarily negative in the limit  $a \to 0$, 
which demonstrates that the full Hamiltonian can become arbitrarily negative and the kinetic energy piece does not offset this divergence.

\subsection{Smooth configuration}

A disadvantage of the above configuration for $B_0$ is that it is
discontinuous, which makes it difficult to implement it in a numerical
simulation. 
We can use the following continuous
configuration, instead of Eq.\ \eqref{initial_B0_stepwise},
employing a linear combination of Gaussians:
\begin{align}
B_0(r) &= b + f(r), \nonumber \\
f(r) &= \delta\bar{B}_0 e^{-r^2/a^2} 
+ \delta\tilde{B}_0\left(e^{-(r-R)^2/c^2} + e^{-(r+R)^2/c^2}\right)\nonumber\\
 &= \delta\bar{B}_0 e^{-r^2/a^2} 
+ 2 \delta\tilde{B}_0 e^{-(r^2 + R^2)/c^2}\cosh\left(\frac{2rR}{c^2}\right).
\label{initial_B0_continuous}
\end{align}
Here $a$, $c$ represent the width of the Gaussians;
they are separated by a distance $R \gg a,c$.
Note that $f(r) = f(-r)$, so $f(r)$ is an even function of $r$.
The constraint equation \rf{constraint} yields
\begin{align}
\pi_r(r) &= \frac{2\lambda}{r^2}\int_0^r B_0(x)\bigl(B_0(x)^2 - b^2\bigr) x^2dx \nonumber\\
&= \frac{2\lambda}{r^2}\int_0^rf(x)\bigl(b + f(x)\bigr)\bigl(2b + f(x)\bigr)x^2 dx.
\end{align}
which produces a linear combination of error functions divided by $r^2$.
note that, for small $r$, $\pi_r(r)$ is proportional to $r$
(just like in the piecewise-constant case considered above).

Our objective is to obtain a local configuration.
In this case, we demand that the fields $f(r)$ and $\pi_r(r)$ tend to
zero exponentially for large $r$. 
This is automatically the case for $f(r)$.
In order that it also be the case for $\pi_r(r)$,
the total effective charge of the configuration has to vanish
(just like the piecewise-constant construction).
For this to be the case we need to impose the condition
\begin{equation}
\int_0^\infty f(x)\bigl(b + f(x)\bigr)\bigl(2b + f(x)\bigr)x^2 dx = 0.
\label{condition_zero-charge}
\end{equation}
This yields a cubic expression in terms of the constants $\delta\bar{B}_0$
and $\delta\tilde{B}_0$,
which can be used to solve $\delta\tilde{B}_0$ in terms of $\delta\bar{B}_0$.

In the limit in which $a,c \ll R$
and $\delta\bar{B}_0,\delta\tilde{B}_0 \gg b$, it follows that
\begin{equation}
\delta\tilde{B}_0 \approx -\frac{a}{\sqrt[3]{3cR^2}}\,\delta\bar{B}_0.
\end{equation}
It is straightforward to show that, in this limit,
the energy of the electric field becomes, at leading order
\begin{equation}
2\pi \int_0^\infty \pi_r^2 r^2 dr =
\frac{\pi^{3/2}\lambda^2(\delta\bar{B}_0\,a)^6}{\sqrt{2\times3^5}} \frac{1}{a},
\label{electric-field-energy_smooth}
\end{equation}
instead of Eq.\ \eqref{electric-field-energy},
and the potential energy is
\begin{equation}
\int_0^\infty \left(\frac{\lambda}{2}(\delta X)^2 -  2\lambda B_0^2\delta X\right)4\pi r^2dr
\approx 
-\frac{3\pi^{3/2}\lambda(\delta\bar{B}_0\,a)^4}{16}\frac{1}{a},
\label{potential-energy-divergent_smooth}
\end{equation}
instead of Eq.\ \eqref{potential-energy-divergent}.
The sum of Eqs.\ \eqref{electric-field-energy_smooth} and
\eqref{potential-energy-divergent_smooth} diverges to $-\infty$ in the limit
$a \to 0$, 
while maintaining $\delta\bar{B}_0 a$ constant,
if
\begin{equation}
\bigl(\delta\bar{B}_0\,a\bigr)^2 < \left(\frac{3}{2}\right)^{7/2}\frac{1}{\lambda},
\end{equation}
which is the equivalent of the condition \eqref{condition_Q}.

\section{Study of field equations with spontaneous Lorentz-symmetry breaking}
\label{QFT}

In the following we will concentrate on the stable bumblebee model described
in the previous section, with the cubic potential
$V = -\frac{\lambda}{3}(X - x)^3$.
We will assume $x > 0$, amounting to timelike expectation value for
the bumblebee field.
The time derivatives of the field components follow by taking
the Poisson brackets with the primary Hamiltonian,
the space-integral of Eq.\ \eqref{primary-Hamiltonian},
where the coefficient $\mu$ is given by Eq.\ \eqref{mu}.
It follows that \cite{Escobar:2022lpu}
\begin{align}
\label{dot-Bi}
\dot{B}_i &= \pi_i + \partial_i B_0, \\
\label{dot-Pi}
\dot{\pi}_i &= \partial_j B_{ji} + 2V' B_i, \\
\label{dot-B0}
\dot{B}_0 &= \frac{\partial_i(V' B_i) + 2B_0 B_i (\pi_i + \partial_i B_0)V''}{V' + 2B_0^2 V''},\\
\label{ivc}
\partial_i\pi_i &= 2B_0 V'(X),
\end{align}
where the last equation is the initial value constraint equation.

\subsection{Expansion around the vev}
\label{expansion around the vev}

Let us consider small fluctuations around a constant vacuum expectation value:
\begin{equation}
B_\mu = b_\mu + \beta_\mu,
\label{ssb_exp}
\end{equation}
where we will consider $b_\mu$ on the minimum of the potential, i.e.,
\begin{equation}
b_0^2 - b_ib_i - x = 0.
\end{equation}
It follows that 
\begin{align}
X &= (b_0 + \beta_0)^2 - (b_i + \beta_i)(b_i + \beta_i) \nonumber\\
&= \bar{X} + \delta X,
\end{align}
with
\begin{align}
\bar{X} &= b_0^2 - b_i b_i = x, \\
\delta X &= 2(b_0 \beta_0 - b_i \beta_i) + \beta_0^2 - \beta_i \beta_i.
\label{delta-X}
\end{align}
The cubic potential then takes the form
\begin{equation}
V(X) = -\frac{\lambda}{3}(X - x)^3 = -\frac{\lambda}{3} (\delta X)^3,
\label{cubic_potential}
\end{equation}
and for its derivatives
\begin{equation}
V'(X) = -\lambda (\delta X)^2,\qquad
V''(X) = -2\lambda\, \delta X.
\end{equation}
From Eqs.\ \eqref{dot-Bi}-\eqref{dot-B0} we find the Hamilton equations become
\begin{align}
\label{dot-betai}
\dot{\beta}_i &= \pi_i + \partial_i \beta_0, \\
\dot{\pi}_i &= \partial_j \beta_{ji} - 2\lambda B_i(\delta X)^2  \label{dot-pii}\\
&= \partial_j \beta_{ji} + \mathcal{O}(\beta^2) 
\label{dot-pii-linear},\\
\dot{\beta}_0 &= 
\frac{2 B_i\bigl(B_0(2\partial_i \beta_0 + \pi_i) - B_j\partial_i \beta_j\bigr) 
+ \frac{1}{2}(\delta X)\partial_i \beta_i}
{2B_0^2 + \frac{1}{2}\delta X} \label{dot-beta0} \\
          &= \frac{b_i}{b_0^2}\bigl(2b_0 \partial_i \beta_0
- b_j\partial_i \beta_j + b_0 \pi_i\bigr) + \mathcal{O}(\beta^2).
\label{dot-beta0-linear}
\end{align}
(Note that $B_0 = b_0 + \beta_0$, $B_j = b_j + \beta_j$,
$\delta X$ is given by Eq.\ \eqref{delta-X} and $\beta_{ij}=\prt_i \beta_j -\prt_j \beta_i$.)
In Eq.\ \eqref{dot-B0}, the numerator and the denominator vanish in the
limit $\delta X \to 0$, but with finite ratio,
leading to the well-behaved, $\lambda$-independent expression \eqref{dot-beta0}.
Remember that Eqs.\ \eqref{dot-betai}--\eqref{dot-beta0-linear} are supplemented
by the initial condition Eq.\ \rf{ivc}.
 
Now suppose that we have a field configuration at the minimum of the potential
at some initial moment $t = 0$.
In that case we have at $t = 0$ the conditions
\begin{align}
    \delta X &= 0, \\
    \tfrac{1}{2}\partial_i\delta X &= B_0\partial_i \beta_0 - B_j\partial_i \beta_j = 0.
    \label{condition-minimum}
\end{align}
Using these identities it is straightforward to verify from Eqs.\ \eqref{dot-betai}
and \eqref{dot-beta0} that $d (\delta X)/dt = 0$.
Thus it follows that,
if initially the configuration is at the minimum of {$\mathcal{H}_V$ at all points of space,
it will stay there.
This parallels a result first obtained in Ref.\ \cite{Bluhm:2008yt}
for a quadratic potential (see also \cite{Escobar:2017fdi}).
It was also shown in Ref.\ \cite{Bluhm:2008yt} that the permanence of the condition $\delta X = 0$
also holds in the presence of a minimal coupling to a conserved current $J^\mu$.

\subsection{Momentum space methods}
\label{momentum space methods}

Let us now determine wave-like solutions of the linearized equations of motion
\eqref{dot-betai}, \eqref{dot-pii-linear} and \eqref{dot-beta0-linear},
in which $\beta_\mu$ and $\pi_i$ are proportional to $e^{ik\cdot x}$.
Some algebra leads to the equations
\begin{align}
\label{eq_1}
k_i \beta_0 - k_0 \beta_i - i \pi_i &=0, \\
\label{eq_2}
\vec{k}^2\,\beta_i - k_i(\vec{k}\cdot\vec{\beta}) + ik_0 \pi_i &= 0, \\
\bigl(k_0 b_0 - 2\vec{k}\cdot\vec{b}\bigr)\beta_0 +
\frac{1}{b_0}\bigl(\vec{b}\cdot\vec k\bigr)\bigl(\vec{b}\cdot\vec \beta\bigr)
+ i\vec{b}\cdot\vec\pi &= 0,
\label{eq_3}
\end{align}
which can be expressed in the matrix form
\begin{equation}
\begin{pmatrix}
k_1 & -k_0 & 0 & 0 & -i & 0 & 0 \\
k_2 & 0 & -k_0 & 0 & 0 & -i & 0 \\
k_3 & 0 & 0 & -k_0 & 0 & 0 & -i \\
0 & k_2^2+k_3^2 & -k_1 k_2 & -k_1 k_3 & ik_0 & 0 & 0 \\
0 & -k_1 k_2 & k_1^2+k_3^2 & -k_2 k_3 & 0 & ik_0 & 0 \\
0 & -k_1 k_3 & -k_2k_3 & k_1^2+k_2^2 & 0 & 0 & ik_0 \\
k_0b_0-2\vec{k}\cdot\vec{b} & \frac{\vec{b}\cdot k}{b_0}b_1 & \frac{\vec{b}\cdot k}{b_0}b_2 & \frac{\vec{b}\cdot k}{b_0}b_3 & ib_1 &  ib_2 &  ib_3
\end{pmatrix}
\begin{pmatrix}
\beta_0 \\ \beta_1 \\\beta_2 \\ \beta_3 \\ \pi_1 \\ \pi_2 \\ \pi_3
\end{pmatrix}
= 0.
\label{matrix_eom}
\end{equation}
First of all, the matrix form \eqref{matrix_eom} implies vanishing determinant
for the square matrix, yielding
\begin{equation}
\frac{ik_0}{b_0}(b\cdot k)^2(k\cdot k)^2 = 0.
\end{equation}
Let us first assume that $k_0 \ne 0$.
Then it follows that either $b\cdot k = 0$
or $k\cdot k = 0$.

Next, by taking the sum of Eq.\ \eqref{eq_2} and $k_0$ times Eq.\ \eqref{eq_1}
it follows that
\begin{equation}
(k\cdot k)\vec{\beta} = (k\cdot \beta)\vec{k},
\end{equation}
This equation admits two types of solutions, namely
either $k\cdot k = k\cdot \beta = 0$ (massless, transverse), or
$k\cdot k \ne 0$ and $\beta_\mu \propto k_\mu$ (longitudinal).
Finally, taking the sum of Eq.\ \eqref{eq_3} and
$\vec{b}\cdot{}$Eq.\ \eqref{eq_1}, it follows that
\begin{equation}
(b\cdot b)(k\cdot b) = 0,
\end{equation}
from which we conclude that either $b\cdot b = 0$ or $k\cdot b = 0$.

Combining these three conditions it follows that there are the following two types
of solutions:
\begin{enumerate}
\item $k\cdot k = 0$ with $\beta \cdot k = \beta \cdot b  = 0$: massless, transverse solutions,
as photons in standard Maxwell theory,
satisfying the additional "gauge condition" $b\cdot \beta = 0$.
These correspond to two local degrees of freedom.
\item $b\cdot k = 0$ with $\beta_\mu \propto k_\mu$: longitudinal solutions, also satisfying $\beta \cdot b = 0$.
In Maxwell theory these modes are gauge artifacts.
They correspond to one local degree of freedom.
In the observer frame in which $b_\mu$ is purely timelike,
they are constant (i.e., zero time derivative).
\end{enumerate}
Note that all these solutions satisfy $\beta \cdot b = 0$, so that $\delta X = 0$.
The fluctuations therefore remain on the minimum of the potential,
and can be all interpreted as Nambu-Goldstone modes of the broken Lorentz symmetry \cite{Bluhm:2004ep}.

Finally, let us consider the remaining case $k_0 = 0$.
Eqs.\ \eqref{eq_1}-\eqref{eq_3} would then appear to yield
an additional nontrivial solution.
However, in this case it is important to remember that the
phase space is restricted by the constraint $\Omega_2 = 0$,
which yields, at linear order in $\beta_\mu$,
the condition $\vec{k}\cdot\vec{\pi} = 0$. 
Combining this condition with Eq.\ \eqref{eq_1} then gives
$\beta_0 = 0 = \vec{\pi}$.
This reduces the solution to a special case of the second (longitudinal)
types of solutions identified above.

We can obtain the mode solutions also directly from the configuration
space equations of motion in Fourier space.
From the Lagrangian density \eqref{Lagrangian} we find:
\begin{equation}
-(k\cdot k)\beta^\mu + k^\mu (\beta \cdot k) - 2V'(X)(b^\mu + \beta^\mu) = 0.
\label{eom}
\end{equation}
Expanding, as above,
\begin{equation}
V'(X) = V'(\bar{X}) + \delta X\,V''(\bar{X}) + \frac{1}{2}(\delta X)^2\,V'''(\bar{X}),
\end{equation}
we see that, at the minimum of the potential where $ V'(\bar{X}) =  V''(\bar{X}) = 0$,
that the last term in Eq.\ \eqref{eom} is quadratic in $\beta^\mu$
(note that $\delta X \approx 2b\cdot \beta$).
Therefore, at linear order, Eqs.\ \eqref{eom} are identical to those of unperturbed Maxwell theory. 
We discuss the Maxwell theory correspondence in the general nonlineanr case in the next section.

These equations (\rf{eom}) can be satisfied in the first place by taking $\beta^\mu \not\propto k^\mu$, 
in which case we find $k\cdot k = \beta \cdot k = 0$,
while at quadratic order we need to impose, additionally, the condition
$\delta X = b\cdot \beta  = 0$.
These solutions exactly correspond to the transverse massless photon solutions
encountered above.
The second possibility are longitudinal mode solutions, $\beta^\mu \propto k^\mu$,
such that the first two terms in Eq.\ \eqref{eom} cancel.
Again, we need to impose, additionally, the condition
$\delta X = 2 b\cdot \beta = 0$.

\section{Solutions of the classical field equations}
\label{solutions of the classical field equations}

We now study the classical field equations in either the Hamiltonian form given
in Eqs.\ \rf{dot-Bi}-\rf{ivc}
or the Euler-Lagrange form.
The latter can be obtained from the original action with variation with respect to $B_\mu$,
yielding
\begin{equation}
    \partial_\nu(\partial^\nu B_\mu - \partial_\mu B^\nu) = 2 V'(X) B_\mu.
    \label{eom-covariant}
\end{equation}
Eqs.\ \eqref{eom-covariant} can be derived from the Hamiltonian equations of motion Eqs.\ \rf{dot-Bi}-\rf{dot-B0} together with the constraint \rf{ivc}.
Explicitly, the $\mu = i$ equation follows by suitably combining the time derivative of Eq.\ \eqref{dot-Bi} and Eq.\ \eqref{dot-Pi}, while the $\mu = 0$ equation follows from the divergence of Eq.\ \eqref{dot-Bi} and the constraint \eqref{ivc}. 
Also, note that by taking the four-divergence $\prt^\mu$
of both sides of Eq.\ \eqref{eom-covariant} one obtains
\beq
\prt^\mu (2 V'(X) B_\mu )=0.
\label{current-cons}
\eeq
This equation is equivalent to the Hamiltonian equation of motion \eqref{dot-B0}; this can be checked by solving the equation for $\dot {B}_0$}.

\subsection{Solutions at the Lorentz-breaking minimum}

It was noted in the previous section that in momentum space, at linear order in fluctuations around $X = x$, the equations of motion of the bumblebee reduce to those of Maxwell theory. 
There is another way the correspondence to Maxwell theory can be seen, using the observation in the previous section that if the condition $X = x$ (and thus $V'(X) = V''(X) = 0$) is satisfied initially
at all points of space, it is guaranteed to be satisfied at all times.
This result was obtained first in Ref.\ \cite{Bluhm:2008yt}.
Consequently, for these solutions the bumblebee equations of motion \eqref{eom-covariant} simply reduce to those of Maxwell theory,
with the added feature that they satisfy $X = B_\mu B^\mu = x$.
In other words, they correspond to those of Maxwell theory in the nonlinear gauge $B_\mu B^\mu = x$.
The bumblebee with smooth potential, with the initial configuration satisfying $X = x$, is also equivalent to the Nambu model \cite{Nambu:1968qk,Kostelecky:2003fs,Escobar:2017fdi}.
\\

In the next subsections we will consider solutions that deviate from the condition $X = x$.

\subsection{Static solutions, spherical symmetry}
\label{static, spherical}

In this section we will investigate some static solutions of the equations
of motion.
In particular, we will focus on spherically symmetric solutions (in Minkowski spacetime).
Taking the static, spherically symmetric
ansatz $B_\mu = \bigl(B_0(0,r,0,0),B_r(0,r,0,0),0,0\bigr)$,
Eqs.\ \eqref{eom-covariant} become
\begin{align}
    \label{eom-B_0}
    -\frac{1}{r^2}\frac{\partial}{\partial r}\left(r^2\frac{\partial B_0}{\partial r}\right) &= 2B_0\,V'(X), \\
    0 &= 2B_r\,V'(X),
    \label{eom-B_r}
\end{align}
where $X = B_0^2 - B_r^2$.
From Eq.\ \eqref{eom-B_r} we see that either $B_r = 0$ or $V'(X) = 0$.

Let us first investigate the possibility $B_r = 0$.
In that case Eq.\ \eqref{eom-B_0} reduces to the nonlinear second-order
differential equation
\begin{equation}
    -\frac{1}{r^2}\frac{\partial}{\partial r}\left(r^2\frac{\partial B_0}{\partial r}\right)
    + 2\lambda B_0(B_0^2 - x)^2 = 0,
    \label{eom-B_0_B_r=0}
\end{equation}
Writing $B_0(r) = \sqrt{x} + \beta(r)$ (we assume the timelike case $x>0$),
we have, in the limit $\beta \ll \sqrt{x}$, the approximate equation
\begin{equation}
    \frac{1}{r^2}\frac{\partial}{\partial r}\left(r^2\frac{\partial \beta}{\partial r}\right)
    - A \beta^2 = 0,
    \label{eom-b}
\end{equation}
with $A = 8\lambda x^{3/2}$.
By properly rescaling the variables,
Eq.\ \eqref{eom-b} can be cast into the form of the Lane-Emden equation
\begin{equation}
    \frac{1}{\xi^2}\frac{\partial}{\partial \xi}\left(\xi^2\frac{\partial \theta}{\partial \xi}\right)
    + \theta^n = 0,
    \label{Lane-Emden}
\end{equation}
(with $\xi$ and $\theta$ dimensionless), for $n = 2$.
This equation is well known in the literature and has been used,
for various values of $n$,
to investigate the interior of stars.
In that case, one generally integrates Eq.\ \eqref{Lane-Emden} numerically
starting at $\xi = 0$, with initial conditions $\theta(0) = 1$, $\theta'(0) = 0$.
As it turns out, a recursive power series solution is known for the case $n = 2$,
with a radius of convergence that is at least 1.

In our case, we do not necessarily need to start to integrate at $r = 0$.
Let us try, instead, the ansatz
\begin{equation}
    \beta(r) = \beta_0 r^m.
\end{equation}
Substituting into Eq.\ \eqref{eom-b} yields
\begin{equation}
    m(m+1)\beta_0 r^{m-2} = A \beta_0^2 r^{2m},
\end{equation}
from which we find $m = -2$ and $\beta_0 = 2/A$.
Thus
\begin{equation}
    B_0 = \sqrt{x} + \frac{2}{Ar^2},
    \label{B_0-approx}
\end{equation}
which is valid as long as the second term is small,
i.e.,  $Ar^2/2 \ll \sqrt{x}$.
Clearly this differs from the standard Maxwell electrostatic behavior $B_0 \sim 1/r$.

\begin{figure}[h!]
    \includegraphics[width=0.70\linewidth]{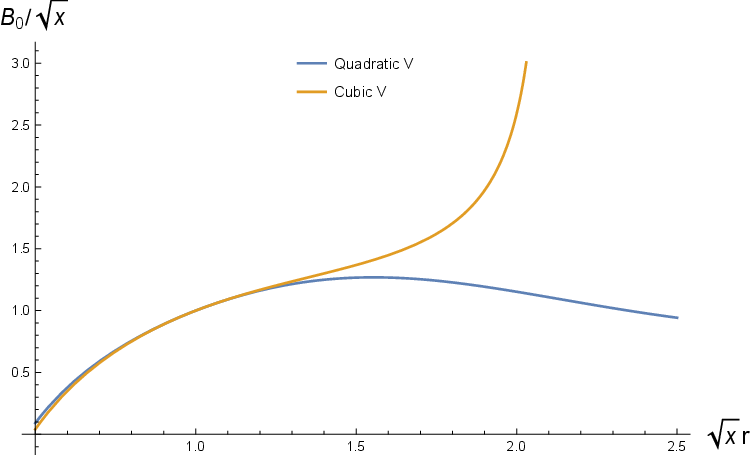}
    \caption{A plot of the static solution away from the potential minimum for the cubic potential and the quadratic potential.  The cubic curve continues to diverge.  The boundary conditions are $B_0=1$ and  $\prt_{\sqrt{x} r} (B_0/\sqrt{x})=1$ at $\sqrt{x} r=1$.}
    \label{fig:staticplot}
 \end{figure}

For small $r$, the approximation \eqref{B_0-approx} breaks down,
and we need to consider the full differential equation \eqref{eom-B_0_B_r=0}.
By introducing $u = r^{-1}$ it can be written as 
\begin{equation}
    u^4\frac{d^2 B_0}{du^2} = 2\lambda B_0(B_0^2 - x)^2.
    \label{u-eqn}
\end{equation}
For positive $B_0$,
the right-hand side of this equation is semi-positive definite,
so that the second derivative of $B_0$ is always nonnegative,
and, therefore, the graph of $B_0$ is always concave.
As a consequence, in the limit $u \to \infty$ (i.e., $r \to 0$),
the continuation of the solution given by Eq.\ \eqref{B_0-approx}
will diverge to infinity.
In general, the latter observation implies that there can be no nontrivial static
solutions that are finite in the full range $0 \le r < \infty$.
To illustrate this point we show here sample solutions for $B_0$ with the cubic potential versus the quadratic potential
in figure \ref{fig:staticplot}.
In this limited time-independent context, 
the static solution for quadratic potential was shown to be stable in Ref.\ \cite{bmw25}.
This figure is the solution to \rf{eom-B_0} with $B_r=0$ where we plot $B_t/\sqrt{x}$ and $\sqrt{x} r$.
One can see the consequences of the concavity argument above as the solution for the cubic potential is unstable
in this static scenario.
We return to this apparent instability when we discuss time dependence later.

Next, consider the second possibility of solving Eq.\ \eqref{eom-B_r},
namely $V'(X) = 0$, implying $B_0^2 - B_r^2 - x = 0$.
This fixes $B_r$ algebraically as a function of $B_0$:
\begin{equation}
    B_r = \pm \sqrt{B_0^2 - x}.
    \label{eom-algebraic}
\end{equation}
Eq.\ \eqref{eom-B_0} now turns into the \textit{linear} differential
equation
\begin{equation}
     -\frac{1}{r^2}\frac{\partial}{\partial r}\left(r^2\frac{\partial B_0}{\partial r}\right) = 0,
     \label{eom-B_0-linear}
\end{equation}
which is solved by
\begin{equation}
    B_0(r) = \alpha + \frac{\beta}{r},
    \label{sol-B_0}
\end{equation}
for arbitrary constants $\alpha$ and $\beta$.
It then follows from Eq.\ \eqref{eom-algebraic} that
\begin{equation}
    B_r(r) = \pm\sqrt{\left(\alpha + \frac{\beta}{r}\right)^2 - x}.
    \label{sol-B_r}
\end{equation}
Eq.\ \eqref{sol-B_r} always produces a real solution as long as we choose
$\alpha^2 - x \ge 0$.

This second type of static solution represents the equivalent of the electric field
created by a point charge in Maxwell theory.
Interestingly, it involves nonzero configurations of both the time and
the radial components of the vector field,
as opposed to the field configuration in Maxwell theory,
where only the time component is relevant.
(In Maxwell theory, the presence of the radial component of the form
\eqref{sol-B_r} would represent a gauge artifact.)
This interpretation makes it clear that
the solution described by Eqs.\ \eqref{sol-B_0} and \eqref{sol-B_r}
arises only in the presence of a point charge at $r = 0$,
which is the only point where the solution \eqref{sol-B_0} diverges and does not
strictly satisfy the equation of motion \eqref{eom-B_0-linear}.
Consequently, the existence of this type of solution implies a coupling
to an external charged current density. 

\subsection{Other Special Cases}
\label{other special cases}

In this subsection we take another approach: investigate solutions to the equations of motion with some simplifications.
We look at homogeneous solutions 
(spatially invariant) but retaining time-dependent solutions.
Then we investigate static solutions but only spatial dependence along one direction, 
in contrast to the previous spherical symmetry assumptions above.

\subsubsection{Time dependence only}
\label{time dependence only}

In Cartesian coordinates, 
we assume the form $B_\mu = (B_0(t), B_1(t),B_2(t),B_3(t))$.
This choice may seem esoteric but is the case for the homogeneity of space.
Since each component is only time dependent, Eqs.\ \eqref{eom-covariant} reduce to: 
\begin{align}
\label{Cartesian Time}
\ddot{B}_i&=2V'(X)B_i, \\
         0&=2V'(X) B_0.
\label{timeeqns}
\end{align}
where either $B_0=0$ or $V'(X)=0$.
In the case where $V'(X)=0$, 
the solution for $B_i$ is trivial:  
\begin{equation}
   B_i=C_{1i}t+C_{2i},
   \label{Bisoln1}
\end{equation}
where $C_{1i}$ and $C_{2i}$ are constants.
We can then solve for $B_t$ using $V'(X)=0$, 
which implies $X=B_0^2-B_1^2-B_y^2-B_z^2=x$.
Thus we have,
\begin{equation}
    B_0(t)=\pm \sqrt{x +\sum_{i=1}^3(C_{1_i}t+C_{2_i})^2},
    \label{Btsoln1}
\end{equation}
for which the timelike case ($x>0$) is guaranteed to be real-valued for all $t$.
For the spacelike case
there could be times $t$
when the square argument could become negative since $x<0$. However,
the bumblebee components are real-valued for all $t$ even for $x<0$, 
if $C_{1,1} = C_{1,2} = C_{1,3} = 0$, i.e., no time dependence at all. 
The field then takes a constant, uniform value on the stationary hypersurface.
Note that for generic values of the $C_{1i}$, 
the solutions \rf{Bisoln1} and \rf{Btsoln1} grow in magnitude without limit, 
and are unlikely to be physically relevant on long time scales.
However, they are independent of the choice of potential $V$.

\begin{figure}[h!]
 \begin{subfigure}[h]{0.48\linewidth}
   \includegraphics[width=\linewidth]{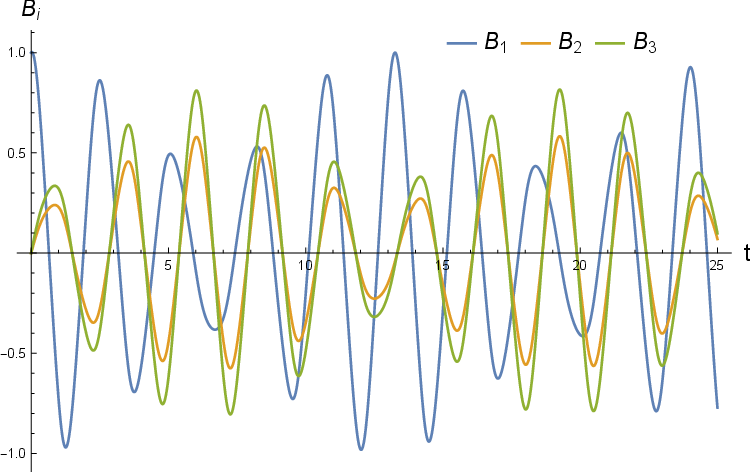}
   \caption{A set of numerical solutions for Eqs.\ \eqref{Bieqns} for the case of $x>0$ (timelike).  The initial values of $B_i$ can be gleaned from the graph, while the initial derivatives are chosen as $\dot{B}_1=0.2$,$\dot{B}_2=0.5$, and $\dot{B}_3=0.7$.}
   \label{timelike-x}
 \end{subfigure}
\hfill
 \begin{subfigure}[h]{0.48\linewidth}
   \includegraphics[width=\linewidth]{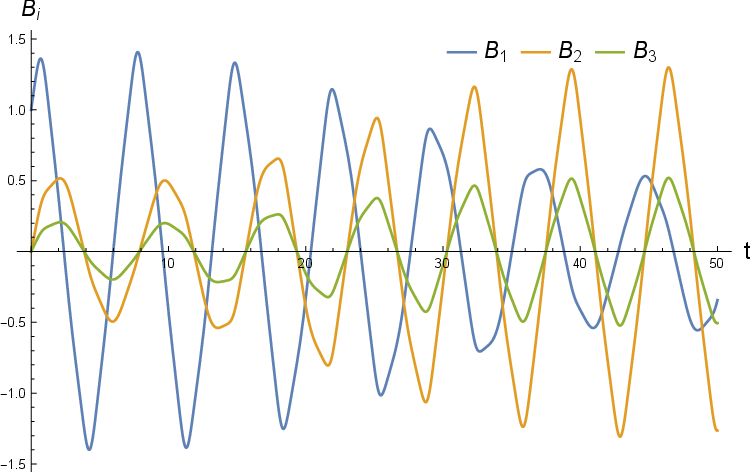}
   \caption{A set of numerical solutions for Eqs.\ \eqref{Bieqns}
    for the case of $x<0$ (spacelike).  The initial values of $B_i$ can be gleaned from the graph, while the initial derivatives are chosen as $\dot{B}_1=0.2$,$\dot{B}_2=0.5$, and $\dot{B}_3=0.7$.}
   \label{spacelike-x}
 \end{subfigure}
\caption{Plots of numerical solutions of Eqs.\ \eqref{Bieqns}.}
\label{TimeDepOnly}
\end{figure}

Alternatively, 
we can set $B_0=0$, 
resulting in a system of three nonlinear ordinary differential equations.
For simplicity here, 
we focus on the cubic potential $V=\lambda (X-x)^3/3$, 
for which the equations become,
\begin{align}
    \frac{d^2 B_i}{dt^2}&=-2\lambda(B_1^2 + B_2^2 + B_3^2 + x)^2 B_i.
\label{Bieqns}
\end{align}
When $x>0$ (timelike extrema), 
there is no real solution to $X=x$, 
thus the solutions do not include the minimum of the potential.
On the other hand, 
when $x<0$ (spacelike extrema), 
the solutions do include the minimum of the potential.
The equations \rf{Bieqns} are challenging to solve analytically, 
but we can numerically evaluate them.
Specifically, 
we display sample solutions for the timelike $x>0$ case in Figure \ref{timelike-x},
and the spacelike $x<0$ case in Fig.\ \ref{spacelike-x}.

In the timelike case, 
the results resemble coupled oscillations with varying amplitudes.
This is not surprising since, 
in the absence of a minimum for the potential, 
the equations resemble an anharmonic oscillator.
For the spacelike case, 
the solutions are periodic, 
albeit with non-sinusoidal oscillations
as seen in the second figure \ref{spacelike-x}.

\begin{figure}[h!]
 \begin{subfigure}[h]{0.48\linewidth}
   \includegraphics[width=\linewidth]{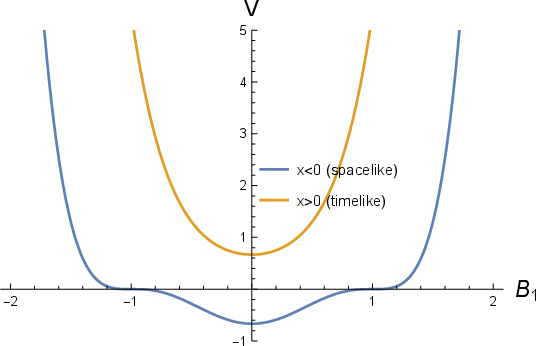}
   \caption{Plot of the potential energy functions from Eq.\ \rf{Bieqns3}.
    The cases of spacelike minima ($x<0$) and timelike minima ($x>0$) are shown.  Units are chosen such that $x=\pm 1$.}
   \label{1Dpotplot}
 \end{subfigure}
\hfill
 \begin{subfigure}[h]{0.48\linewidth}
   \includegraphics[width=\linewidth]{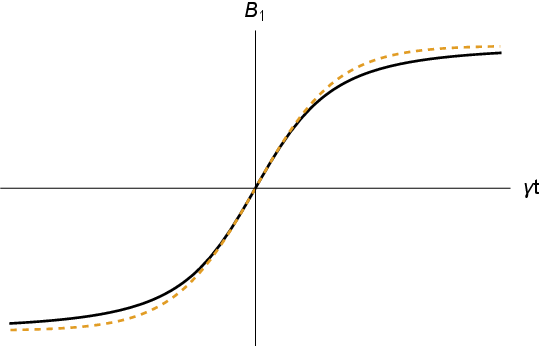}
   \caption{Plot of the temporal kink solution in \rf{C10soln}, solid curve; the dashed curve is a $\tanh \gamma t$ solution for comparison.}
   \label{kink}
 \end{subfigure}%
\caption{Plots of aspects of the solutions to Eq.\ \rf{Bieqns3}.}
\label{Bieqns_plots}
\end{figure}

We next consider a more specialized case where we set two vector components, $B_2$ and $B_3$, equal to zero,
reducing Eqs.\ \rf{Bieqns} to:
\begin{equation}
    \frac{d^2B_1}{dt^2}=-2\lambda B_1(x +B_1^2)^2.
    \label{Bieqns2}
\end{equation}
Since $\frac{d^2B_1}{dt^2}=\dot{B}_1\frac{d\dot{B}_1}{dB_1}$, we can recast the equation as
\begin{equation}
    \bigg(\frac{dB_1}{dt}\bigg)^2=
    -\frac{2\lambda}{3}(x + B^2_1)^3+C_1,
    \label{Bieqns3}
\end{equation}
which can be interpreted as the classical mechanics energy equation with kinetic energy $(dB_1/dt)^2$, 
potential energy $2\lambda (x+B_1^2)^3/3$ and total energy $C_1$.
This reveals that the value of $C_1$ has to be greater than or equal to the minimum of the potential, generically.
A plot of the effective potential energy functions 
$2\lambda (x+B_1^2)^3/3$ is shown in figure \ref{1Dpotplot}.
One can see from energy considerations that the solutions will be periodic and stable for viable values of $C_1$. 

There is a special case solution that exists when $C_1=0$ and $x<0$ (spacelike).
One can think of this as the case where the ``particle" is at rest at the unstable stationary point where the potential energy is zero.
For this case, Eq.\ \rf{Bieqns3} can be integrated yielding
\beq
B_1 (t) = \sqrt{|x|}\left( \frac{\pm \gamma t +t_0}{\sqrt{1+(\pm \gamma t +t_0)^2}} \right),
\label{C10soln}
\eeq
where $\gamma = |x| \sqrt{2\lambda/3}$
and $t_0$ is an arbitrary instant in time.
This solution actually resembles a temporal kink-type solution in which the particle moves from its initial position $\mp \sqrt{|x|}$ at $t\rightarrow -\infty$
to its final position $\pm \sqrt{|x|}$ at $t\rightarrow \infty$.
In terms of the bumblebee field, 
this solution represents a transition from one side of the one-sheeted hyperboloid $X=x$ on which $\cal H_V$ is stationary (but for $x<0$, not at its minimum, as we saw in section \ref{sec:model}) to the other.
We plot this result in figure \ref{kink}.
This type of behavior could be of interest for cosmological studies.
In contrast, for the timelike case $x>0$ the solution for $C_1=0$ is imaginary for some portion of the $x$ axis.

\subsubsection{One dimensional spatial dependence}
\label{one dimensional spatial dependence}

We now consider a static scenario with dependence on just one direction, 
thus $B_\mu=(B_0(x),B_1(x),0,0)$ 
(the more general static scenario was already discussed in Sec.\ \ref{static, spherical}). 
In this limit, 
Eq.\ \eqref{eom-covariant} yields:
\begin{align}
    \label{eom-B_t_carteisan_1D_static}
     \frac{d^2 B_0}{d x^2}&=- 2B_0\,V'(X), \\
    0 &= 2B_1\,V'(X).
    \label{eom-B_1_1D_static}
\end{align}
Like the time-dependent case, 
Eq.\ \eqref{eom-B_1_1D_static} allows us to choose $B_1=0$ or $V'(X)=0$.
If we take $V'(X)=0$, 
we get 
an analogous result to Eqs.\ \eqref{Bisoln1} and \eqref{Btsoln1}:
\begin{align}
    B_0(x)&=C_1x+C_2, \\
    B_1(x)&=\pm\sqrt{(C_1x+C_2)^2- (\pm b^2)},
    \label{x-soln}
\end{align}
where we use $\pm b^2$ in lieu of $x$ to avoid confusion with the coordinate $x$ ($+b^2$ is the timelike minimum case).
If $C_1 \neq 0$,
the magnitudes of these fields simply grow without limit for increasing or decreasing $x$,
If $C_1 =0$, 
the fields have constant values.

For the alternative solution with $B_1=0$ we arrive at a similar equation to \rf{Bieqns3}
but with the role of time and space switched.
Specifically for the case of the cubic potential we have,
\begin{equation}
    \bigg(\frac{dB_0}{dx}\bigg)^2=\frac{2\lambda}{3}(B_0^2 \mp b^2)^3+C_1.
    \label{x-eqn}
\end{equation}
If $C_1=0$ this equation can be solved for 
timelike and spacelike cases to yield, 
respectively, 
\beq
\bal
B_0 (x) &= b\left( \frac{\pm \gamma x +C_2}{\sqrt{(\pm \gamma x +C_2)^2-1}} \right),\\
B_0 (x) &= b\left( \frac{\pm \gamma x +C_2}{ \sqrt{1-(\pm \gamma x +C_2)^2}} \right).
\label{C10solnx}
\eal
\eeq
In this case, 
singular values of $x$ arise when the square root
argument vanishes, 
and a real valued solution does exist for a portion of the real line.
This provides a contrast between the cubic potential, 
and the quadratic potential;
the latter of which was shown to admit domain wall solutions in $x$ \cite{seifert10}.
Other solutions for $C_1 \neq 0$ can be countenanced but are omitted here for brevity.

\subsection{Numerical simulation of field evolution with spherical symmetry}
\label{numerical simulation of the field equations}

\subsubsection{Set up and coordinate choices}
\label{set up}

Let us consider the evolution of configurations with spherical symmetry.
In that case the relevant field components in phase space are
$B_0(t,r) = b_0 + \beta_0(t,r)$, $B_r(t,r) = b_r + \beta_r(t,r)$ and $\pi_r(t,r)$.
Equations of motion \eqref{dot-betai}--\eqref{dot-beta0-linear} and the initial condition \eqref{ivc} become
\begin{align}
    \label{dot-beta_r}
    \dot \beta_r &= \pi_r + \partial_r \beta_0, \\
    \label{dot-pi_r}
    \dot\pi_r &= 2B_rV', \\
    \label{dot-beta0_2}
    \dot \beta_0 &= \frac{B_r\bigl(B_0(2\partial_r \beta_0 + \pi_r) - B_r\partial_r \beta_r\bigr) + \frac{V'}{2V''}\left(\partial_r \beta_r + \frac{2}{r}B_r\right)}{B_0^2 + \frac{V'}{2V''}}, \\
    \label{constraint-pi_r2}
    \frac{1}{r^2}\partial_r\left(r^2 \pi_r\right) &= 2B_0 V', 
\end{align}
where, for the cubic potential, $V' = -\lambda(\delta X)^2$
and $V'/V'' = \frac{1}{2}\delta X$.
In order to check the stability of the system,
we consider an initial configuration with a fluctuation away from the minimum of the potential,
for instance $b_0 = \sqrt{x}$, $b_r = 0$, and
\begin{align}
     \label{beta0-initial}
     \beta_0(0,r) &= \epsilon e^{-a r^2}, \\
     \beta_r(0,r) &= 0, 
     \label{betar-initial}
\end{align}
where $\epsilon$ satisfies $\epsilon \ll \sqrt{x}$.
The function on the right-hand side of Eq.\ \eqref{constraint-pi_r2} then becomes,
at $t = 0$,
\begin{equation}
    -2\lambda B_0(0,r)\bigl(\delta X(0,r)\bigr)^2 =
-8\lambda\epsilon^2x^{3/2}e^{-2ar^2}\left(1 + \frac{2\epsilon}{\sqrt{x}}e^{-ar^2} +
\frac{5\epsilon^2}{4x}e^{-2ar^2} + \frac{\epsilon^3}{4x^{3/2}}e^{-3ar^2}\right).
\label{initial-charge-density}
\end{equation}
Denoting the right-hand side of Eq.\ \eqref{initial-charge-density} as $f(r)$
one obtains from \rf{constraint-pi_r},
\begin{equation}
    \pi_r(0,r) = \frac{1}{r^2}\int_0^r (r')^2f(r') dr',
    \label{pi-r-initial}
\end{equation}
which satisfies $\pi_r(0,0)  = 0$ and, for large $r$, 
$\pi_r(0,r) \propto 1/r^2$.

The evolution equations \eqref{dot-beta_r}--\eqref{dot-beta0_2} can now be used to determine $\beta_\mu(t,r)$ and $\pi_r(t,r)$ for $t > 0$ numerically.
In this case one needs to supplement the evolution equations by boundary
conditions at $r = 0$.
In order that the spatial parts of $\beta_\mu$ and $\pi_\mu$ be non-singular
at the origin,
the radial components $\beta_r$ and $\pi_r$ need to vanish at $r = 0$.
It then follows from Eq.\ \eqref{dot-beta_r} that the radial derivative
of $\beta_0$ vanishes as well.
In other words, we have to impose:
\begin{equation}
    \beta_r(t,0) = \pi_r(t,0) = \partial_r \beta_0(t,0) = 0.
\end{equation}

One problem that arises in Eqs.\ \eqref{dot-beta_r}-\eqref{constraint-pi_r} is the divergent behavior
at $r \to 0$ of Eq.\ \eqref{dot-beta0_2}.
Fortunately, it can be overcome by taking the radial functions
$B_r(t,r)$ and $\pi_r(t,r)$ to be odd functions of $r$,
and $B_0(t,r)$ to be an even function of $r$.
It is straightforward to check that these symmetries
are consistent with Eqs.\ \eqref{dot-beta_r}-\eqref{constraint-pi_r},
and with the initial configuration chosen above.
In this case we have
$b_r = 0$ and thus $\beta_r(r) = B_r(r)$.
We can then define the functions
$\tilde{B}_r = B_r/r$ and
$\tilde{\pi}_r = \pi_r/r$.
As they are symmetric in the variable $r$,
they are analytic functions of $s = r^2$.
This is also the case for $B_0$ and $\beta_0$.
It is straightforward to check that,
in terms of these functions, Eqs.\ \eqref{dot-beta_r}-\eqref{constraint-pi_r}
become
\begin{align}
    \label{dot-tilde-beta_r}
    \dot{\tilde B}_r &= \tilde\pi_r + 2\partial_s \beta_0, \\
    \label{dot-tilde-pi_r}
    \dot{\tilde\pi}_r &= 2 \tilde B_r V', \\
    \label{dot-beta0_3}
    \dot\beta_0 &= \frac{s\tilde B_r\bigl(B_0(4\partial_s \beta_0 + \tilde\pi_r) - \tilde B_r(1 + 2s\partial_s)\tilde B_r\bigr) + \frac{V'}{2V''}\left((1 + 2s\partial_s)\tilde B_r + 2\tilde B_r\right)}{B_0^2 + \frac{V'}{2V''}}, \\
    \label{constraint-tilde-pi_r}
    (3 + 2s\partial_s)\tilde\pi_r &= 2B_0V'.
\end{align}
These equations do not have any more singularity as $s \to 0$.
They can be used to study the evolution of the spherically symmetric configuration but it is challenging to deal with $s\rightarrow \infty$.

To facilitate boundary conditions
in a numerical simulation, 
such as demanding field fluctuations vanish as $r\rightarrow \infty$,  
we next compactify the infinite range of the radial variable to the compact interval $[0,1[$.
This can, for instance, be accomplished by the following transformations:
\begin{align}
    u &= \frac{r^2}{r_0^2 + r^2},\\
    B_r &= \tanh\left(\frac{r}{r_0}\right)\tilde{B}_r, \\
    \pi_r &= \tanh\left(\frac{r}{r_0}\right)\tilde{\pi}_r,
\end{align}
where we can take the transformed functions $\tilde{\beta}_r$, $\tilde{B}_r$ and $\tilde{\pi}_r$ 
to be analytic functions of $u$.
It can be checked that the Hamiltonian equations of motion in the new variables become:
\begin{align}     
    \dot{\tilde{B}}_r &= \tilde{\pi}_r + \frac{f(u)}{r_0 g(u)}\partial_u\beta_0,
    \label{dot-tilde-beta-r-2}\\
    \dot{\tilde{\pi}}_r &= 2\tilde{B}_r V', 
    \label{dot-tilde-pi-r-2}\\
    \dot\beta_0 &= 
    \Biggl\{\tilde B_rB_0\bigl[2f(u)\partial_u\beta_0 +  r_0 g(u)\tilde\pi_r\bigr]
   - g(u)\tilde B_r^2\bigl(1 - g(u) + f(u)\partial_u\bigr)\tilde B_r
   \nonumber\\
     & \qquad{} 
     + \frac{V'}{2V''}\left[\bigl(1 - g(u) + f(u)\partial_u\bigr)\tilde B_r + \frac{f(u)}{u(1-u)}\tilde B_r\right]\Biggr\}\frac{1}{r_0\left(B_0^2 + \tfrac{V'}{2V''}\right)},
    \label{dot-beta-0-2}
 \end{align}
and the constraint equation is
\beq
         \left( 1-g +
          \frac {f(u)}{u(1-u)} 
          + f(u)\partial_u
          \right)
          \tilde {\pi}_r = 2 B_0 V',
\label{Pi_constraint}
\eeq
where we define, 
\begin{align}
    f(u) &= 2u^{1/2}(1-u)^{3/2}\tanh\left(\sqrt{\frac{u}{1-u}}\right), \\
    g(u) &= \left(\tanh\left(\sqrt{\frac{u}{1-u}}\right)\right)^2.
\end{align}
The evolution equations \eqref{dot-tilde-beta-r-2}--\eqref{dot-beta-0-2} are analytic in $u$ and non-divergent on the interval $0 \le u < 1$.

\subsubsection{Hyperbolicity of the field equations}
\label{hyperbolicity}

A positive Hamiltonian does not guarantee a well-posed initial value problem for the bumblebee field.
We examine the hyperbolicity conditions on the spherically-symmetry restricted field equations in \rf{dot-beta_r}-\rf{constraint-pi_r}.
\footnote{Note that an analysis of hyperbolicity was performed for other vector field models in \cite{sarbach19}.}
These equations can be cast in the form
\begin{align}
    \dot \beta_r - \partial_r \beta_0&= \pi_r , \\
    \dot\pi_r &= 2B_rV', \\
    \dot \beta_0 - f\partial_r\beta_0 - g\partial_r\beta_r &= h
\end{align}
where
\begin{align}
f &= \frac{2B_rB_0}{B_0^2 + \frac{V'}{2V''}},\\
g &= \frac{\frac{V'}{2V''} - B_r^2}{B_0^2 + \frac{V'}{2V''}},\\
h &= \frac{B_0B_r\pi_r + \frac{V'}{rV''}B_r}{B_0^2 + \frac{V'}{2V''}}.
\end{align}
We can write this in vector form:
\begin{equation}
\begin{bmatrix}\dot\beta_r\\ \dot\pi_r\\ \dot\beta_0\end{bmatrix} +
\begin{bmatrix}
0 & 0 & -1\\
0 & 0 & 0\\
-g & 0 & -f
\end{bmatrix}
\begin{bmatrix}\partial_r\beta_r\\ \partial_r\pi_r\\ \partial_r\beta_0\end{bmatrix} =
\begin{bmatrix} \pi_r\\ 2B_rV'\\ h \end{bmatrix}
\label{PDE-vector-form_3}
\end{equation}
Now note that in our case, the variables $\{\beta_r,\pi_r,\beta_0\}$ are not independent,
but constrained by the condition
\begin{equation}
    \label{constraint-pi_r}
    \frac{1}{r^2}\partial_r\left(r^2 \pi_r\right) = 2B_0 V'.
\end{equation}
In particular, this means that $\pi_r$ can be expressed in terms of $B_0$ and $B_r$.
Consequently, we can eliminate it without any problem from the system \eqref{PDE-vector-form_3},
which then reduces to
\begin{equation}
\begin{bmatrix}\dot\beta_r\\ \dot\beta_0\end{bmatrix} +
\begin{bmatrix}
0 & -1\\
-g & -f
\end{bmatrix}
\begin{bmatrix}\partial_r\beta_r\\ \partial_r\beta_0\end{bmatrix} =
\begin{bmatrix} \pi_r\\ h \end{bmatrix}
\label{PDE-vector-form_2}
\end{equation}
together with the constraint \eqref{constraint-pi_r}.

Hyperbolicity is determined by the characteristic matrix in the second term of Eq.\ \eqref{PDE-vector-form_2}.
Generally, the system is hyperbolic if this matrix has only real eigenvalues, and is diagonalizable.
The eigenvalues are given by
\begin{equation}
\lambda_\pm = \frac{-f \pm \sqrt{f^2 + 4g}}{2}
\end{equation}
The latter are real iff $f^2 + 4g \ge 0$, which is equivalent to the condition
\begin{equation}
\frac{V'}{2V''}\left(B_0^2 - B_r^2 + \frac{V'}{2V''}\right) = 
\frac{V'}{2V''}\left(X + \frac{V'}{2V''}\right) \ge 0
\end{equation}
For the cubic potential $V'/V'' = \frac{1}{2}(X-x) = \frac{1}{2}\delta X$,
so that this condition reduces to
\begin{equation}
\delta X\left(\frac{5}{4} \delta X + x\right) \ge 0,
\label{hyperbolic}
\end{equation}
which implies (assuming $x > 0$) that $\delta X \ge 0$ or $\delta X \le -\frac{4}{5}x$. 
The eigenvectors corresponding to $\lambda_+$ and $\lambda_-$ are given by
\begin{equation}
 \begin{bmatrix}
1\\ -\lambda_+
\end{bmatrix}
\qquad \text{and}\qquad
\begin{bmatrix}
1\\ -\lambda_-
\end{bmatrix}
\end{equation}
respectively.
If $f^2 + 4g > 0$, $\lambda_+ \ne\lambda_-$, and thus the two eigenvectors are independent,
and the characteristic matrix diagonalizable.
On the other hand, if $f^2 + 4g = 0$
(which happens, in particular, on the minimum of the effective Hamiltonian $\mathcal{H}_V$),
$\lambda_+ = \lambda_-$, and therefore there is only one independent eigenvector.
The system becomes then weakly hyperbolic at that specific state,
being at the boundary between hyperbolic (wave-like) and parabolic (diffusion-like).
The consequence is that the solution can experience algebraic growth over time.
Instead of variables splitting into independent waves,
terms multiplied by time ($t \partial_r u$) appear in the analytical solution.

Interestingly, note that for the case $\delta X = 0$ the (single) eigenvector is
\begin{equation}
\begin{bmatrix}
1 \\ -B_r/B_0
\end{bmatrix}.
\end{equation}
This corresponds exactly to the direction parallel to the minimum-$\mathcal{H}_V$ surface $X = x$.
Therefore, if the initial condition is taken to be on this surface, 
the evolution will be hyperbolic (with the configuration remaining on the surface).
As it turns out, the condition \eqref{hyperbolic} actually defines the hyperbolic regime for the cubic potential in general, not just for spherically symmetric configurations.
See Appendix \ref{ap:hyperbolic} for a proof.

\subsubsection{Numerical Simulation of bumblebee evolution}
\label{simulation}

\begin{figure*}
  \centering
     \subfloat[Time and space evolution of the spherically symmetric solution for $B_0$ and $B_r$ for the cubic potential; we choose $a=2$ for the Gaussian spread.]
     {
         \includegraphics[width=0.48\textwidth]{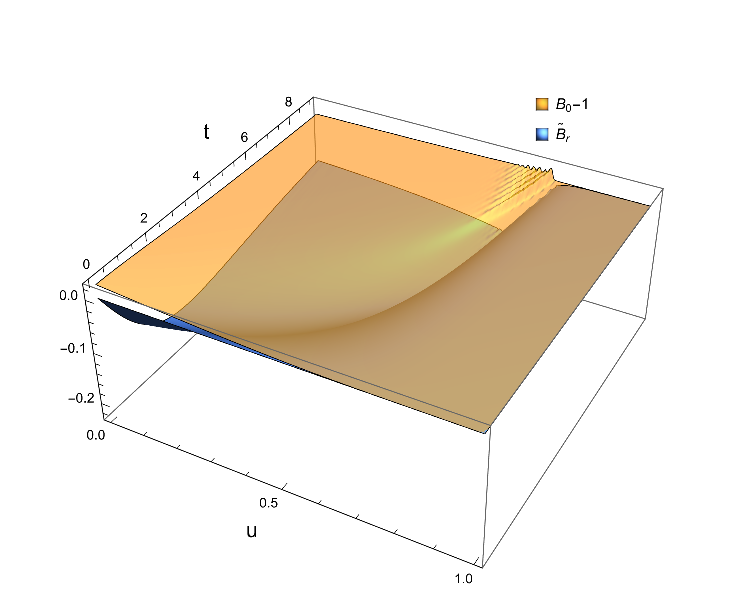}
         \label{fig:cubic1}}
     \hfill
     \subfloat[Time and space evolution of the argument of the potential $X-x$ for the cubic potential with $a=2$.]{
         \includegraphics[width=0.48\textwidth]{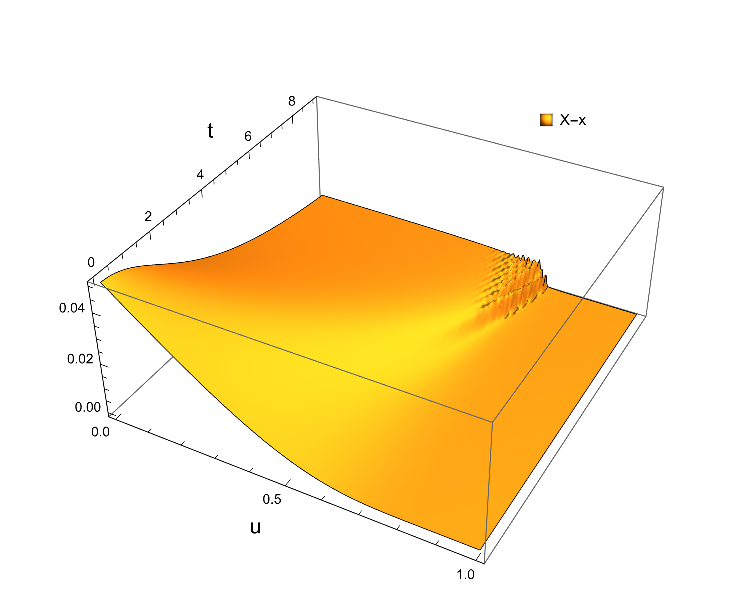}
         \label{fig:cubic2}} 
         \\
     \subfloat[Time and space evolution of the spherically symmetric solution for $B_0$ and $\tilde B_r$ using an initial Gaussian function with a wider spread $a=1/2$.]{
        \includegraphics[width=0.48\textwidth]{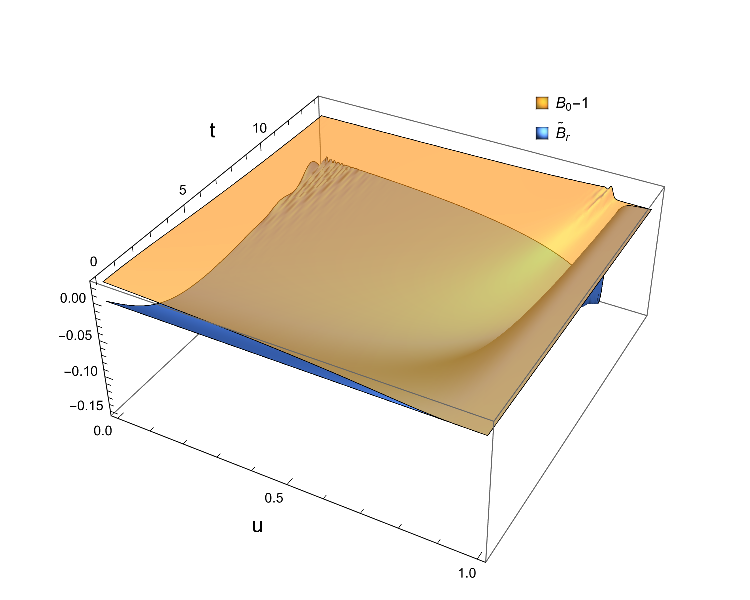}
         \label{fig:cubic3}} 
      \hfill
      \subfloat[Time and space evolution of the argument of the potential $V$ using an initial Gaussian function with a wider spread $a=1/2$.]{
         \includegraphics[width=0.48\textwidth]{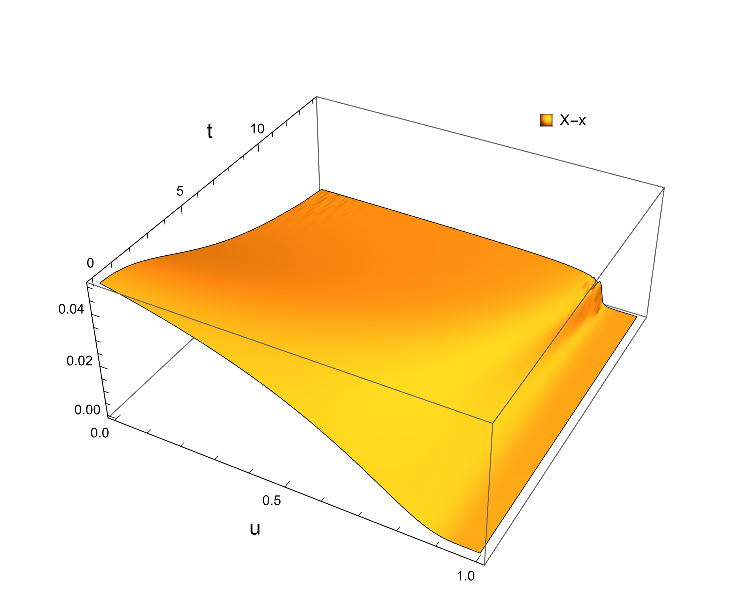}
         \label{fig:cubic4}} 
      \caption{Numerical solutions of the field equations \rf{dot-tilde-beta-r-2}-\rf{dot-beta-0-2} with initial configuration \rf{beta0-initial} and \rf{betar-initial}.  Units are chosen such that $x=1$.  The size of the Gaussian initial perturbation is given by $\epsilon=1/40$.  In all these figures the development of shock behavior can be clearly observed.}
       \label{numplots}
\end{figure*}

Using the equations \rf{dot-tilde-beta-r-2}-\rf{dot-beta-0-2}, 
the initial value constraint equation
\rf{Pi_constraint},
and the initial pertubation in \rf{beta0-initial}-\rf{betar-initial} we examine the evolution of the fields using a numerical solver in Mathematica.
In order that the initial value problem is well posed, it is important that the initial configuration is in the hyperbolic regime.
The resulting time evolutions are plotted in Figs.\ \ref{numplots}.
Figs. \ref{fig:cubic1} and \ref{fig:cubic2} are based a Gaussian perturbation with width $a=2$, 
while Figs. \ref{fig:cubic3} and \ref{fig:cubic4} are based on a wider initial Guassian with $a=1/2$. 
In these plots, 
there are several important aspects to highlight.  
Firstly, 
consider Figs.\ \ref{fig:cubic1} and \ref{fig:cubic3}, 
which show the evolution of the bumblebee field components for the cubic potential.
The initial Gaussian in $B_0$ ``seeds" a large response from the $B_r$ term.
Secondly, 
this leads to shock front formation in $B_0$ and $B_r$.
Further, 
we see the same behavior in Figs.\ \ref{fig:cubic2} and \ref{fig:cubic4}, 
that the fluctuation away from the vacuum solution $\delta X$ solutions also form a shock.

\begin{figure}[h]
\includegraphics[width=0.70\linewidth]{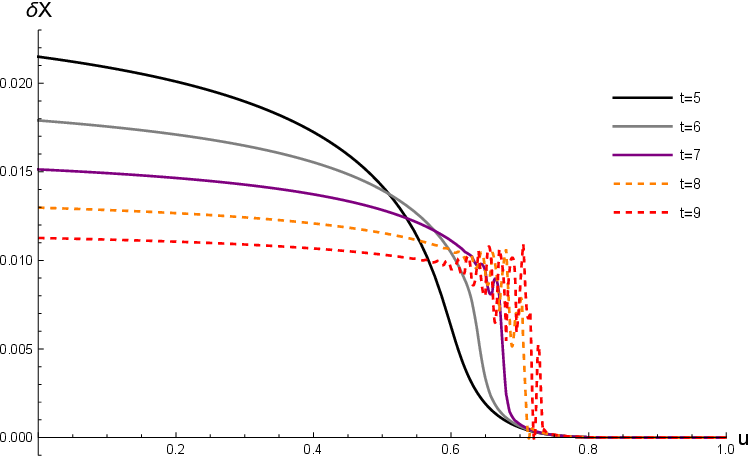}
    \caption{Snapshots in time from the plot Figure \ref{fig:cubic2}.  The dashed curves indicate that the solution samples points out of the hyperbolic regime.  These results are for the cubic potential.}
    \label{fig:shock}
\end{figure}

\begin{figure}[h]
\includegraphics[width=0.70\linewidth]{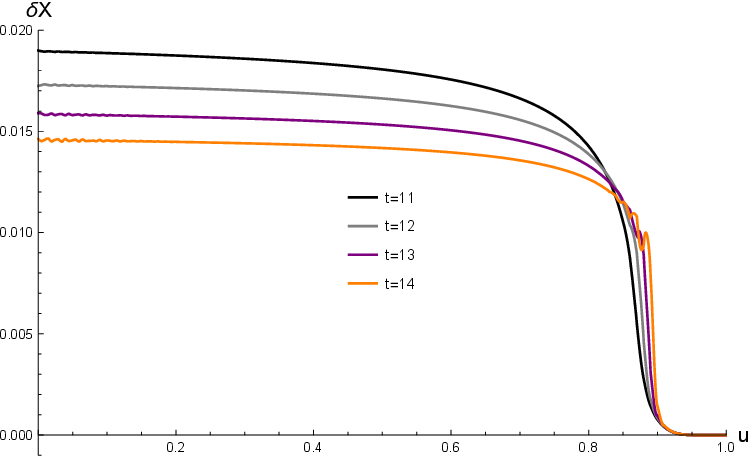}
    \caption{Snapshots in time from the plot Figure \ref{fig:cubic4}. For this case, all of the samples lie within the hyperbolic evolution regime.  The small oscillations near $u=0$ are a numerical artifact.}
    \label{fig:shock2}
\end{figure}

To elucidate the point further, 
the evolution from the initial Gaussian to the shock behavior
is plotted in a series of snapshots of time in Figures \ref{fig:shock} and
\ref{fig:shock2}.
When the evolution reaches a certain point, 
not long after the shock formation, 
the nature of the PDEs can change from hyperbolic to elliptical, 
thus the time evolution of the system breaks down at this point.
In Fig.\ \ref{fig:shock}, 
two sample times where the values of the field dip out of the hyperbolic regime,
given by condition \eqref{hyperbolic},
are shown with dashed lines.
In contrast, 
when the initial Gaussian width is wider, 
the shock forms later in time while the evolution maintains hyperbolicity for some time, 
as seen in Fig.\ \ref{fig:shock2}.

Note that shock behavior has been shown to be possible in other vector field models of spontaneous spacetime-symmetry breaking 
as discussed in the next subsection. 
Furthermore, 
the equations solved here are nonlinear, 
and it is well known that nonlinear PDEs can lead to shock behavior, 
as in,
for example, 
the well-known Burgers equation in fluid dynamics \cite{burgers48}.

\begin{figure}[h!]
 \begin{subfigure}[h]{0.48\linewidth}
   \includegraphics[width=\linewidth]{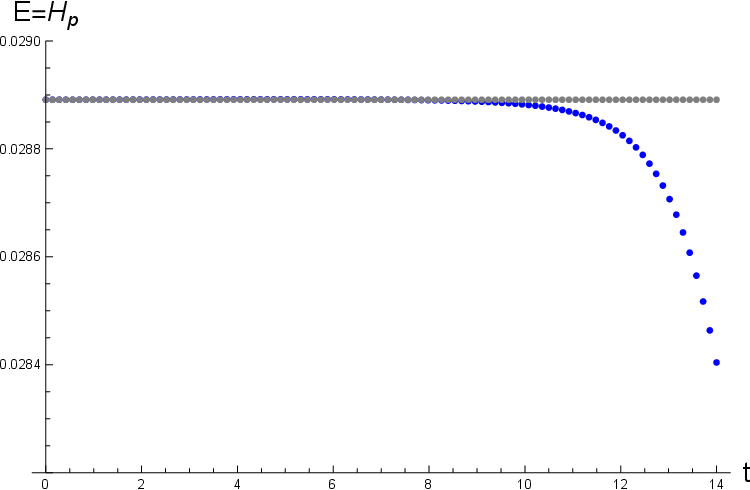}
   \caption{Plot of the total energy of the bumblebee field for the simulation of Figs.\ \ref{fig:cubic3} and \ref{fig:cubic4}.}
   \label{energy}
 \end{subfigure}
\hfill
 \begin{subfigure}[h]{0.48\linewidth}
   \includegraphics[width=\linewidth]{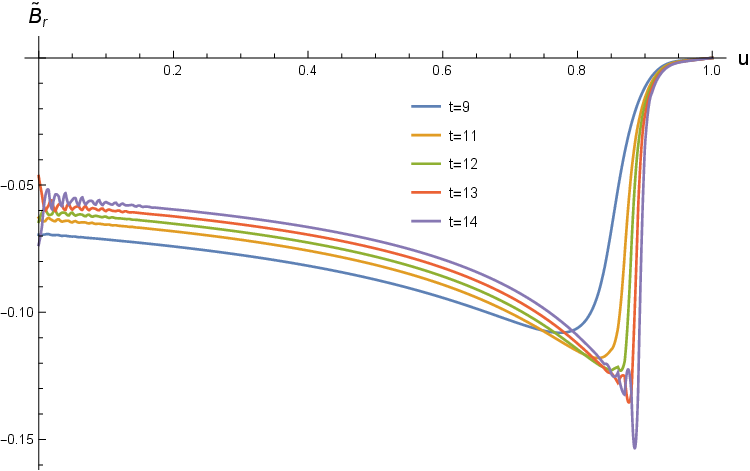}
   \caption{Plot of the radial component ${\tilde B}_r (u)$.}
   \label{Btr}
 \end{subfigure}%
\caption{Plots of the total energy and the radial component $\tilde B_r (u)$.  The total energy is constant up until the shock behavior begins.}
\label{Bieqns_plots}
\end{figure}

Since the dynamics we have explored are based on Hamilton's equations, 
and there are no explicit time-dependent terms in the Hamiltonian, 
the total energy must be conserved.
Using the integral over all space $H_p=\int d^3 x {\cal H}_p$, 
where $\mathcal{H}_p$ is given by Eq.\ \eqref{H_p}, 
we can numerically calculate the total energy for the bumblebee field and check on its constancy.
This is shown in Fig.\ \ref{energy}.
We also show for comparison a plot of snapshots of $\tilde B_r (u)$ approaching the shock behavior in Fig.\ \ref{Btr}.
We can see the energy is constant for much of the simulation.
However, by comparing with the $\tilde B_r$ evolution, 
we see that the total energy ceases to be constant once the shock behavior sets in.
In fact, this is to be expected because the evolution of the field variables becomes ill-defined,
as we will show in the next subsection.

\subsection{Shock discontinuities}
\label{Shock discontinuities}

\begin{figure*}
  \centering
  \includegraphics[width=6cm]{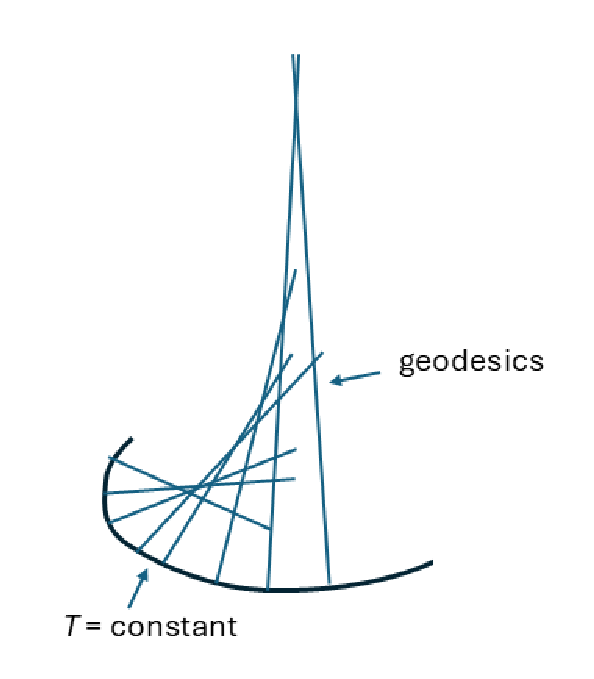}
  \caption{Formation of shocks in the bumblebee. For the gradient form $B_\mu = \partial_\mu T$, $B_\mu$ is tangent to geodesics. If these are launched orthogonal to an initial surface $T = \mbox{constant}$, they will in general cross after some finite proper time, forming shock discontinuities.}
  \label{fig:geodesics}
\end{figure*}

In the previous subsection we saw that, starting with some initial configuration for the bumblebee field, the numerical simulations yield the formation of shock-like discontinuities.
In fact, this should not come as a surprise, because it is in fact a generic feature of the bumblebee with Maxwell kinetic term.
To see this, we use an adaptation of the argument in Ref.\ \cite{Eling:2004dk} showing that the occurrence of shocks, or caustics, is generic in the Einstein-aether model, that is, for the bumblebee coupled to gravity in which the norm is fixed through a Lagrange-multiplier potential.

Consider again the Lagrangian
\begin{equation}
\mathcal{L} = -\frac{1}{4}B_{\mu\nu}B^{\mu\nu} - V(B_\mu B^\mu)
\end{equation} 
where $B_{\mu\nu} = \partial_\mu B_\nu - \partial_\nu B_\mu$,
and $V$ is some potential such the corresponding Hamiltonian
has a minimum at $B_\mu B^\mu = b^2$.
The field equations read
\begin{equation}
\frac{1}{2}\partial_\nu B^{\mu\nu} - 2V'(B_\nu B^\nu)B^\mu = 0.
\label{eom2}
\end{equation}
Let us now restrict ourselves
to configurations that can be written as the gradient of a scalar $T$:
\begin{equation}
B_\mu = \partial_\mu T.
\end{equation}
For these configurations the equations of motion \eqref{eom2} become
\begin{equation}
-2V'\bigl((\partial_\nu T)(\partial^\nu T)\bigr) \partial_\mu T = 0.
\end{equation}
It thus follows that the configuration sits at the bottom of the potential:
\begin{equation}
V'\bigl((\partial_\nu T)(\partial^\nu T)\bigr) = 0,
\end{equation} 
and thus
\begin{equation}
(\partial_\nu T)(\partial^\nu T) = b^2.
\label{condition}
\end{equation}
Then $B_\mu$ satisfies the parallel transport condition
\begin{equation}
B^\nu\partial_\nu B_\mu = (\partial^\nu T)(\partial_\nu\partial_\mu T)
= \frac12 \partial_\mu\bigl((\partial^\nu T)(\partial_\nu T)\bigr) = 0,
\end{equation}
where we used condition \eqref{condition} in the last step.
This means that $B^\mu$ must be tangent to geodesics,
i.e., straight lines (as we consider zero curvature here).
If such geodesics initiate orthogonal to an initial surface of
constant $T$,
they will generically cross after some finite proper time (see Fig.\ \ref{fig:geodesics}).
Where they cross the value of $B^\mu$ is ill-defined,
and the derivative $\partial_\mu B_\nu$ is singular.
These are the shock discontinuities, beyond which the evolution of the bumblebee field cannot be continued. 

\section{Discussion and Conclusion}
\label{discussion}

In this paper, 
the bumblebee model with Maxwell kinetic term and general potential energy function $V(B^\mu B_\mu )$ is studied for its stability and dynamical properties in flat spacetime
and the possibility to induce spontaneous Lorentz breaking,
expanding upon prior work in the literature.

From an analysis of the potential portion of the Hamiltonian \eqref{H_V},
we confirmed and extended recent work \cite{Zhu:2026hxm} that there is a class of potentials $V(X)$ for which the Hamiltonian density of the Bumblebee is bounded below.
Examples include the (shifted) cubic potential $V(X) = -\frac{\lambda}{6}(X-x)^3$,
as well as more general cases of polynomial or non-polynomial form,
such as those given by Eqs.\ \eqref{V-odd}
and \eqref{tanh-cubed-potential}.
We note that the latter provides an example of a potential that can induce spontaneous Lorentz breaking,
even though the Hamiltonian density does not attain its absolute minimum at $X = x$, but only a local one.
 
We have demonstrated in Section \ref{quadraticstudy} 
that the solutions to the equations for the quadratic potential are generically unstable. 
This is done by studying the full Hamiltonian \rf{Ham-onshell} with two different initial values perturbed from the minimum; the resulting perturbations are shown to result in arbitrarily negative energies.
Thus the positive kinetic energy terms cannot 
compensate for the negative potential portion of the Hamiltonian;
as can be seen in Eq.\ \rf{total-divergent}.

Another set of key results is contained in the study of the linear aspects in Sec.\ \ref{QFT}.
Here we discussed the expansion of the bumblebee field $B_\mu$ around its vacuum value $b_\mu$ in Sec.\ 
\ref{expansion around the vev}, 
followed by a detailed study of the momentum space equations of motion in Sec.\ \ref{expansion around the vev}.
The latter includes a characterization of the degrees of freedom of the model with the cubic potential when the equations are linearized.

A comprehensive study of classical solution to the field equations (in flat spacetime) was undertaken in Sec.\ \ref{solutions of the classical field equations}.
We first covered static solutions in Sec.\ \ref{static, spherical}, 
with a study of the spherically-symmetric solutions of Eqs.\ \rf{eom-B_0}-\rf{eom-B_r}.
We found instability of generic static solutions for the cubic potential, as can be seen in 
Figure \ref{fig:staticplot}.
On the other hand, 
well-behaved solutions for $B_\mu$ lying at the minima exist (Eqs.\ \rf{sol-B_0}-\rf{sol-B_r}).
In Sec.\ \ref{other special cases}, 
we considered special case solutions with time-dependence only, or a simplified $B_\mu$ field pointing along $x$ only.
The results include the oscillatory solutions 
of Figs.\ \ref{timelike-x} and \ref{spacelike-x}, 
and the intriguing temporal kink solution in 
Eq.\ \rf{C10soln} and Fig.\ \ref{kink}.

Of particular importance is Section \ref{numerical simulation of the field equations}, 
wherein we simulated a spherically symmetric solution to the bumblebee equations evolving from an initial perturbation around the potential minimum.
We pointed out that the bumblebee model can contain regions of field space where the evolution equations cease to be well-posed. 
In fact, the evolution can transition from hyperbolic to elliptical,
as shown in subsection \ref{hyperbolicity} and Appendix \ref{ap:hyperbolic}.
This is an important feature that arises in the equations due to the nonlinearity of the terms stemming from the potential $V$.

Using advantageous field redefinitions and coordinates, 
the nonlinear coupled PDEs (Eqs.\ \rf{dot-tilde-beta-r-2}-\rf{Pi_constraint}) were solved using computational routines in Mathematica \cite{mathematica}.
The resulting evolution is displayed in Figs.\ \ref{numplots}.
For the cubic potential, 
the results show an initial relaxation of the fields $B_0$ and $B_r$ back to their vacuum values (away from the perturbed initial values).
However, 
we found that after some time, 
shock-like discontinuities occur.
These were further bolstered by an analytic argument using the field equations that such shocks can occur in the bumblebee model (Sec.\ \ref{Shock discontinuities}).

In this paper,
we have focused on the bumblebee model in flat spacetime, 
and explored potentials with positive semi-definite Hamiltonian.
Studying the dynamics we found that shocks and non-hyperbolic evolution can occur in the full nonlinear model.
Adding a coupling to gravity and further studying aspects explored in this paper would be of interest.
For instance, one could examine the evolution of spherically symmetric solutions when coupled to gravity.
Following the identification of the elliptic regime of the bumblebee equations of motion it would be very interesting to develop new methods in order to obtain nontrivial solutions for the field fluctuations around the bumblebee vacuum $X = x$.
In addition, 
the anti-symmetric tensor model of spontaneous spacetime symmetry breaking could also be studied in a similar manner \cite{abk10}.
Note that Hamiltonian issues already exist for these models \cite{seifert19-2,Potting:2023xzt}.

\acknowledgments
RP acknowledges support from
CIDMA--Center for Research and Development in Mathematics and Applications under the Portuguese Foundation for Science and Technology Multi-Annual Financing Program for R\&D Units, grants UID/04106/2025
(https://doi.org/10.54499/UID/04106/2025) and
UID/PRR/04106/2025 (https://doi.org/10.54499/ UID/PRR/04106/2025).
DTWC and QGB acknowledge support for this work from the Undergraduate Research Institute at Embry-Riddle Aeronautical University, Prescott and the Arizona NASA Space Grant consortium.

\appendix
\section{General derivation of the conditions for hyperbolicity}
\label{ap:hyperbolic}

Here we show that condition \eqref{hyperbolic} defines hyperbolicity in general for the cubic potential.
We start from the second-order form of the equations of motion (Eq.\ \eqref{eom-covariant}):
\begin{equation}
\partial_\nu(\partial^\nu B_\mu - \partial_\mu B^\nu) = 2V'(X)B_\mu
\label{eom-bumblebee}
\end{equation} 
It is useful to decompose 
\begin{equation}
B_\mu = B_{\perp,\mu} + B_{\|,\mu}
\end{equation}
such that $B_\perp$ satisfies the (Lorenz-gauge) condition
\begin{equation}
\partial_\nu B_\perp^\nu = 0
\end{equation}
and $B_\|$ is a gradient
\begin{equation}
B_{\|,\mu} = \partial_\mu\tilde\phi.
\end{equation}
Note that $\tilde\phi$ is determined by the equation
\begin{equation}
\square\tilde\phi = \partial_\mu B^\mu.
\end{equation}
Eq.\ \eqref{eom-bumblebee} reduces to
\begin{equation}
\square B_{\perp,\mu} =  2V'(X)(B_{\perp,\mu} + B_{\|,\mu}).
\end{equation}
At first sight this appears to be an explicitly hyperbolic
equation
(note that a system of second-order PDE's is hyperbolic
if the signature of the eigenvalues of the matrix determining the second-order derivatives
is either $(-+++)$ or $(+---)$).
However, the nonlinear term on the right-hand side effectively amounts to a constraint on the field components,
equivalent to the second-class constraints in the Hamiltonian formulation,
that takes the form of a PDE of the scalar field $\phi$.
In order to see this,
we take the divergence
of Eq.\ \eqref{eom-bumblebee} and obtain (Eq.\ \eqref{current-cons})
\begin{equation}
\partial^\mu\bigl(V'(X)B_\mu\bigr) = 0
\label{eom-longitudinal}
\end{equation}
Now defining the shifted scalar field $\phi$ through $\tilde\phi = \phi + b\cdot x$ it follows that
\begin{equation}
B_{\|,\mu} = b_\mu + \partial_\mu\phi
\end{equation}
and
\begin{equation}
X = (\partial_\nu\phi + b_\nu + B_{\perp,\nu})(\partial^\nu\phi + b^\nu + B_\perp^\nu).
\end{equation}
Eq.\ \eqref{eom-longitudinal} becomes
\begin{align}
0 &= V'(X) \partial^\mu B_\mu + B^\mu V''(X) \partial_\mu X \nonumber \\
&= V'(X) \partial^\mu\partial_\mu\phi + B^\mu V''(X) B^\nu\bigl(2\partial_\mu(\partial_\nu\phi + B_{\perp,\nu})\bigr)
\end{align}
which can be expressed as
\begin{equation}
\bigl(V'(X) \eta^{\mu\nu} + 2V''(X)B^\mu B^\nu\bigr)\partial_\mu\partial_\nu\phi
+ \mbox{lower order terms} = 0
\end{equation}
where the lower-order terms do not involve second-order derivatives of $\phi$.
The matrix multiplying the quadratic derivatives of $\phi$ is
\begin{equation}
a^{\mu\nu} = V'(X)\eta^{\mu\nu} + 2V''(X)B^\mu B^\nu.
\end{equation}
Its determinant is
\begin{equation}
\det a = -\bigl(V'(X)\bigr)^4\left(1 + \frac{2V''}{V'}B^\mu B_\mu\right)
\end{equation}
Condition for hyperbolicity is $\det a < 0$, which implies
\begin{equation}
1 + \frac{2V''}{V'} X > 0
\end{equation}
Writing $X = x + \delta X$ we obtain for the cubic potential,
for which $V(X) = -\frac{\lambda}{6}(\delta X)^3$, that the hyperbolic regime is defined by
\begin{equation}
\delta X > 0 \quad\mbox{or}\quad \delta X < -\frac{4}{5}x,
\end{equation}
just as we had found for the special case of spherical symmetry in the Hamiltonian first-order formulation.

\end{document}